\documentclass[useAMS,usenatbib]{mnras}
\pdfoutput=1 

\usepackage{graphicx}
\usepackage{amsmath, amssymb} 
\usepackage{upgreek} 
\usepackage{graphicx, float, subcaption} 
\usepackage{booktabs, array, multirow} 
\usepackage{siunitx} 

\graphicspath{ {./fig/} } 

\hypersetup{
  allcolors = {black}, 
  pdfpagemode = UseNone,
  pdfstartview = FitH
}

\DeclareSIUnit{\jansky}{\text{Jy}}
\DeclareSIUnit{\arcsec}{\text{arcsec}} 
\DeclareSIUnit{\arcmin}{\text{arcmin}} 
\DeclareSIUnit{\sqdegree}{\ensuremath{\text{deg}^2}} 
\DeclareSIUnit{\beam}{\text{beam}}

\let\originalleft\left
\let\originalright\right
\renewcommand{\left}{\mathopen{}\mathclose\bgroup\originalleft}
\renewcommand{\right}{\aftergroup\egroup\originalright}

\newcommand{\Aegean}{\textsc{Aegean}}

\newcommand{\sub}[2]{#1_\mathrm{#2}} 
\newcommand{\subsup}[3]{#1_\mathrm{#2}^\mathrm{#3}} 
\newcommand{\spindex}[2][]{\ensuremath{\subsup{\alpha}{#2}{#1}}} 

\newcommand{\roughly}[1]{\(\sim\)#1} 
\newcommand{\percentage}[1]{#1~per cent} 

\newcommand{\function}[1]{\texttt{#1}} 
\newcommand{\package}[1]{\textsc{#1}} 

\begin{document}

\title[Characterising broadband radio SEDs]{A novel approach for characterising broadband radio spectral energy distributions}
\author[Harvey et al.]{V.~M.~Harvey$^{1}$\thanks{Email: \href{mailto:violet.harvey@adelaide.edu.au}{violet.harvey@adelaide.edu.au}}, T.~Franzen$^{2,3}$, J.~Morgan$^{2}$, and N.~Seymour$^{2}$ \\
$^{1}$High Energy Astrophysics Group, The University of Adelaide, Adelaide SA 5005, Australia\\
$^{2}$International Centre for Radio Astronomy Research, Curtin University, Bentley WA 6102, Australia\\
$^{3}$CSIRO Astronomy and Space Science, 26 Dick Perry Avenue, Kensington WA 6151, Australia}

\date{Submitted: 6 February 2018}

\pagerange{\pageref{firstpage}--\pageref{lastpage}} \pubyear{2018}

\maketitle

\label{firstpage}

\begin{abstract} 
We present a new broadband radio frequency catalogue across 0.12\,GHz $\le \nu \le$ 20\,GHz created by combining data from the Murchison Widefield Array Commissioning Survey, the Australia Telescope 20~Gigahertz survey, and the literature. Our catalogue consists of 1285~sources limited by $S_{20\,\rm{GHz}} >$ 40\,mJy at $5\sigma$, and contains flux density measurements (or estimates) and uncertainties at 0.074, 0.080, 0.119, 0.150, 0.180, 0.408, 0.843, 1.4, 4.8, 8.6, and 20\,GHz. We fit a second-order polynomial in log-log space to the spectral energy distributions of all these sources in order to characterise their broadband emission. For the 994 sources which are well-described by a linear or quadratic model we present a new diagnostic plot arranging sources by the linear and curvature terms. We demonstrate the advantages of such a plot over the traditional radio colour-colour diagram. We also present astrophysical descriptions of the sources found in each segment of this new parameter space and discuss the utility of these plots in the upcoming era of large area, deep, broadband radio surveys.

\end{abstract}


\section{Introduction}

The radio regime of the electromagnetic spectrum is unique in providing around four orders of magnitude of frequency coverage ($\SI{0.01}{\giga\hertz} \lesssim \nu \lesssim \SI{100}{\giga\hertz}$) from the ground. For technological reasons, large area radio surveys have typically focused on one frequency regime. For example the early radio surveys were conducted at low radio frequencies, \roughly\SI{0.1}{\giga\hertz}, due to the simplicity of the low frequency receiver technology \citep[e.g. 3CR][]{Bennett:62}. Later on, large surveys focused on higher frequencies, particularly around \SI{1.42}{\giga\hertz} \citep[e.g.][]{condon1998} covering the hyperfine transition of the ground state of neutral hydrogen. Relatively few studies have attempted to combine surveys from these separate ranges and consider the broad frequency character of radio sources.

Traditionally, the spectral energy distribution (SED) of radio sources has been represented by radio spectral indices, $\alpha$, defined as $S_\nu \propto \nu^\alpha$. The typical value of $\alpha$ for extra-galactic radio sources is around $-0.7$ \citep{Conway:63} due to synchrotron emission.
The radio spectral index $\alpha$ obeys the relation $\alpha=(\gamma -1)/2$ where $\gamma$ is the spectral index of the energy distribution of the relativistic electrons producing the synchrotron spectrum.

It was realised in the early years of radio astronomy that radio sources were quite complex and presented a wide variety of spectral shapes \citep{Williams:63,Kellermann:69}. When sources show a change in spectral index with frequency this is likely due to a change in the initial electron energy distribution caused by synchrotron or inverse Compton losses. Typically, `extended' sources have $\alpha < -0.5$ from synchrotron emission and `compact' sources have $\alpha \sim 0$ due to the superposition of many components, each with a different low-frequency cutoff due to varying ages and synchrotron self-absorption. Sources showing a clear peak around \SI{1}{\giga\hertz} are known as gigahertz-peaked spectrum (GPS) sources \citep{Conway:63}, and are thought to be young and extremely compact (angular size less than \SI{0.1}{\arcsec}) with a spectral turnover caused by strong synchrotron self-absorption or, in some cases, possibly by free-free absorption \citep{Callingham:15}.

Star-forming galaxies (SFGs) have also been shown to display complex radio SEDs. It has long been known that SFGs comprise both a steep synchrotron component and a flat component from free-free emission from \ion{H}{ii} regions \citep{Condon:92}. The ratio of synchrotron to free-free luminosity is typically \roughly\percentage{10} at \SI{1.4}{\giga\hertz}, although it can vary as shown by \cite{Galvin:16} who ascribe the variation to differing starburst ages. \cite{Clemens:10} have shown a wide variety of SEDs for SFGs showing turnovers at low frequency. These turnovers are likely due to free-free absorption given the large number of \ion{H}{ii} regions in SFGs, but the amount of absorption likely depends on the precise geometry of the starforming regions and the viewing angle of the observer. Furthermore, \cite{Clemens:10} have shown evidence for multiple components with free-free turnovers at different frequencies, again likely related to the geometry of the starforming regions on different scales.

As a spectral index from two well-spaced photometric measurements is equivalent to the logarithm of the flux density ratio, two-point spectral indices are equivalent to colour-colour plots commonly used at other frequencies. This radio colour-colour diagram was first developed by \cite{kesteven1977}. Typically such studies require large samples of galaxies with multiple flux density measurements \citep[e.g.][]{AT20Gsource}, or are used to study well-resolved individual galaxies \citep[e.g.][]{Rudnick:94}. However, such methods cannot precisely deal with more than four photometric measurements and cannot always encompass the true range of properties (as we shall demonstrate). Furthermore, the spectral properties of statistically-complete source samples may change depending on the selection frequency and flux density \citep[e.g.][]{Franzen:14}.

Given the large number of radio sources with multi-frequency photometry it is timely to move beyond simple spectral index plots which do not account for the finer detail seen in radio SEDs. In this paper we present a new technique to characterise radio SEDs, particularly those over a wider frequency range and showing deviations from a simple power-law. We primarily use the radio surveys described below.

The Australia Telescope 20~Gigahertz survey \citep[AT20G;][]{AT20Gsource} is a radio survey of the entire southern sky conducted by the Australia Telescope Compact Array over the course of 2004 to 2008. It contains 5890~sources with peak flux densities greater than \SI{40}{\milli\jansky} at \SI{20}{\giga\hertz}. Follow-up observations were carried out for most sources at \SI{4.8}{\giga\hertz} and \SI{8.6}{\giga\hertz} (hereafter `\SI{5}{\giga\hertz}' and `\SI{8}{\giga\hertz}'), and the survey has a resolution on the scale of arcseconds. The Murchison Widefield Array Commissioning Survey (MWACS) is a radio survey of approximately one-quarter of the southern sky conducted by the Murchison Widefield Array in Western Australia towards the end of 2012. The survey was completed in two separate declination scans. The three observed frequency ranges are centred on \SIlist{119; 150; 180}{\mega\hertz}, and have resolutions ranging from \SIrange{6}{3}{\arcmin}.

This paper is organized as follows. In Section~\ref{sec:cross-matching} we present the construction of a catalogue combining data from \SIrange{0.12}{20}{\giga\hertz}. In Section~\ref{sec:modelling} we present the simple SED modelling. In Section~\ref{sec:alpha-phi-intro} we present our new diagnostic diagram based on this fitting. In Section~\ref{sec:catalogue} we ascribe physical interpretations to our new diagnostic plot and present the final catalogue (radio photometry and SED model). We discuss the strengths and limitations of our approach in Section~\ref{sec:discussion} and conclude in Section~\ref{sec:conclusion}.

\section{Cross-matching AT20G with MWACS}
\label{sec:cross-matching}

\subsection{MWACS-3\texorpdfstring{$\bsigma$}{sigma} detection around AT20G sources}
\label{MWACS 3 sigma detections around AT20G sources}

We have expanded the AT20G catalogue to include MWACS flux densities at \SIlist{119; 150; 180}{\mega\hertz}. The region of overlap between AT20G and MWACS is defined as $21^{\mathrm{h}}20^{\mathrm{m}} < \alpha < 08^{\mathrm{h}}00^{\mathrm{m}}$ and $\SI{-58}{\degree} < \delta < \SI{-19}{\degree}$. The total number of AT20G sources lying within this region of sky, which covers an area of \SI{5760}{\sqdegree}, is 1285. We chose not to use the MWACS 5$\sigma$ catalogue by \cite{hurley-walker2014} but instead repeated the source finding using a lower 3$\sigma$ threshold around each of the 1285 AT20G sources in order to maximise the number of detected MWACS counterparts to AT20G sources. The resulting collection of sources was designated `MWACS-3$\sigma$'.

The MWACS data consist of six images, one for each of the three MWACS frequencies at two declination scans centred at $\SI{-27}{\degree}$ and $\SI{-47}{\degree}$. A simple image-plane combination of the data from the two declination scans was not possible because of the very different states of the ionosphere, as noted by \cite{hurley-walker2014}. The source finder \Aegean{} \citep{hancock2012} was run separately on each image. The \textsc{bane} extension of \Aegean{} was used to generate maps of the local RMS noise. The detection threshold was set to $3\sigma$ and only circular regions \SI{15}{\arcmin} in radius were searched around each source. For each detection, an `island' of significant pixels, consisting of contiguous pixels above $2\sigma$, was created. In cases where a source appeared in the region of overlap between the two declination scans, it was considered to only exist in the scan where it was furthest from the edge.

\subsection{Cross-matching}
\label{Cross-matching}

\subsubsection{Monte Carlo simulations to determine optimum search radius}

\begin{figure}
  \includegraphics[scale=0.55]{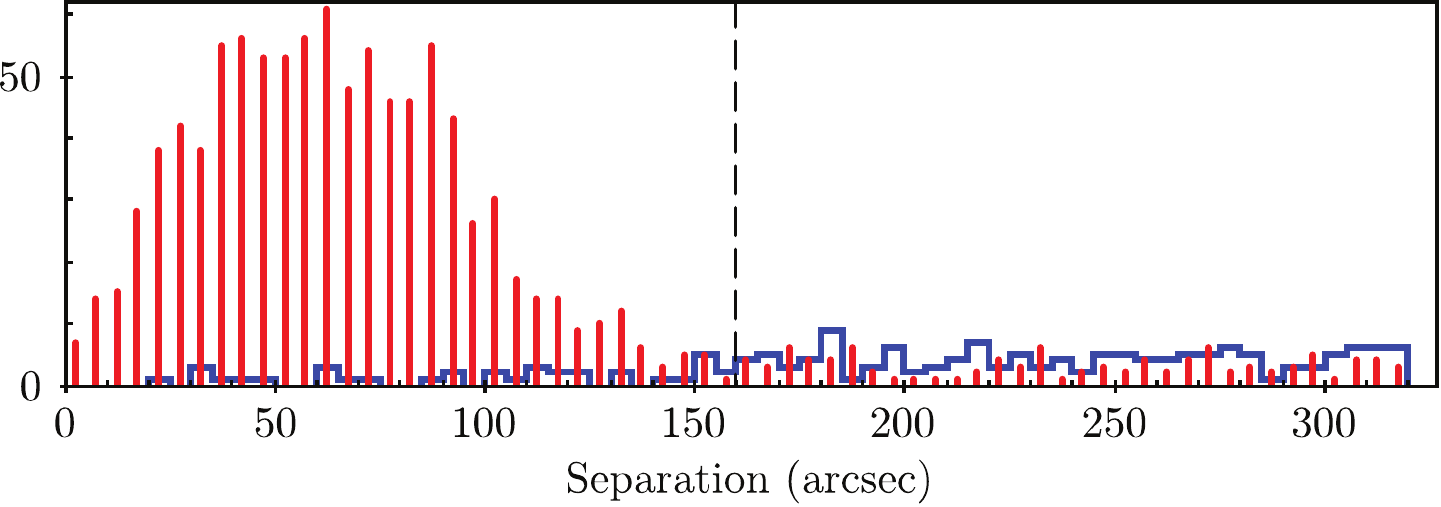}\\
  \includegraphics[scale=0.55]{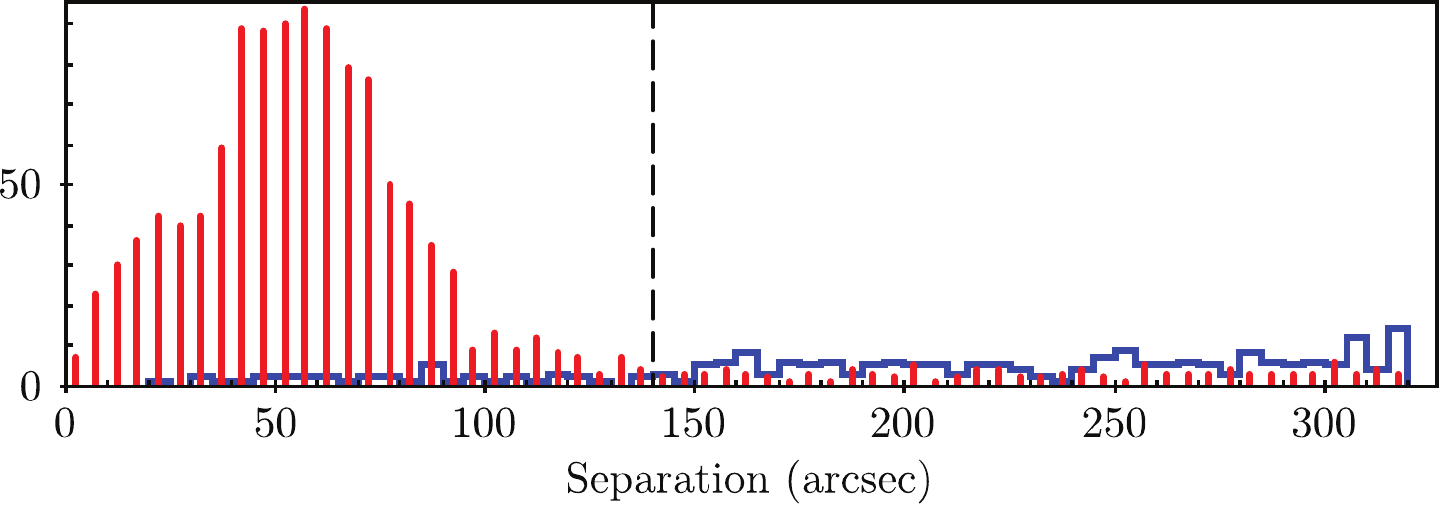}\\
  \includegraphics[scale=0.55]{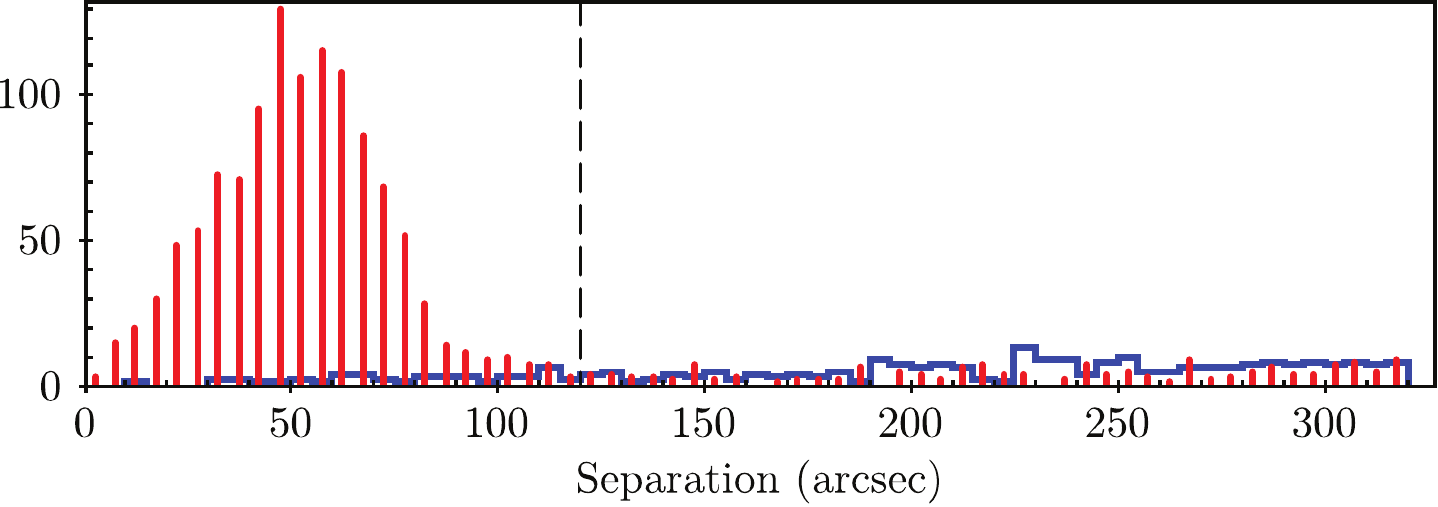}\\
  \caption{Monte Carlo simulations to determine optimum cutoffs for automatic acceptance of AT20G-to-MWACS \SI{119}{\mega\hertz} matches (top panel), AT20G-to-MWACS \SI{150}{\mega\hertz} matches (middle panel) and AT20G-to-MWACS \SI{180}{\mega\hertz} matches (bottom panel). In each case, the red spikes show the number of AT20G sources with identified counterparts in MWACS-3$\sigma$ as a function of angular separation between the AT20G and MWACS-3$\sigma$ positions. The blue steps show the results obtained when cross-matching the MWACS-3$\sigma$ catalogue with a simulated catalogue created by offsetting the AT20G positions by \SI{20}{\degree} in RA and Dec. The dashed vertical line marks the chosen search radius.}
  \label{fig:mc_sim}
\end{figure}

We matched the resulting MWACS-3$\sigma$ source list at each frequency with our AT20G sample, using Monte Carlo simulations to determine the best search radius to automatically accept AT20G-to-MWACS identifications. The results of the simulations are shown in Fig.~\ref{fig:mc_sim}. The red spikes show the number of AT20G sources with identified MWACS counterparts in different bins of angular separation between the AT20G and MWACS positions. A list of simulated sources was created by taking the list of AT20G sources and offsetting their positions by \SI{20}{\degree} in both right ascension and declination. A process identical to that described in Section~\ref{MWACS 3 sigma detections around AT20G sources} was then followed to detect sources above 3$\sigma$ in the MWACS images around these simulated sources. The blue steps show the number of matches between this simulated catalogue and the MWACS catalogue as a function of angular separation.

At \SIlist{119; 150; 180}{\mega\hertz}, the numbers of real and simulated matches per angular separation bin are equal at approximately \SIlist{160; 140; 120}{\arcsec}, respectively. These are taken to be the optimum search radii at each frequency, under the assumption that any matches further out are overwhelmingly likely to be spurious.
The reliability and completeness of the matches at each frequency are estimated in Section~\ref{Reliability and completeness}.

\subsubsection{Iterative catalogue-building}
\label{Iterative catalogue-building}

Since the \SI{180}{\mega\hertz} images have the highest angular resolution and sensitivity, this MWACS frequency was used as the basis for all the AT20G and MWACS-3$\sigma$ matches, before considering \SIlist{150; 119}{\mega\hertz} data. The \SI{180}{\mega\hertz} MWACS-3$\sigma$ catalogue was matched with the AT20G catalogue, and all matches within \SI{120}{\arcsec} were accepted.

Those AT20G sources which had \SI{180}{\mega\hertz} MWACS-3$\sigma$ counterparts were then matched with the \SI{150}{\mega\hertz} MWACS-3$\sigma$ catalogue. All matches within \SI{140}{\arcsec} were accepted. Finally, those AT20G sources which had both \SIlist{180; 150}{\mega\hertz} MWACS-3$\sigma$ counterparts were matched with the \SI{119}{\mega\hertz} MWACS-3$\sigma$ catalogue. All matches within \SI{160}{\arcsec} were accepted. Table~\ref{tab:matchstats} lists the number of AT20G sources with each possible combination of MWACS frequency matches.

We recorded the island integrated flux densities, as measured by \Aegean{}, of all MWACS-3$\sigma$ counterparts to AT20G sources. The uncertainty on the integrated flux density was taken as the sum in quadrature of the local RMS noise and \percentage{10} of the integrated flux density. The \percentage{10} integrated flux density reflected the uncertainty in the absolute flux density scale at these low frequencies and was an important addition for very bright sources as the local RMS would otherwise be too small to represent a reasonable uncertainty. Our overall method for uncertainty estimation had no particular physical motivation, simply being applied because we required some order of uncertainty on the measurements to facilitate the fitting procedure described below, and this was empirically determined to be adequate.

In cases where an AT20G source had no MWACS-3$\sigma$ counterpart at some frequency, an estimate of the true flux density for that frequency was obtained using the value of the nearest pixel to the AT20G source in the corresponding MWACS image, with an uncertainty equal to the local RMS noise. We also used these nearest-pixel values to estimate the flux densities of sources that were confused in the MWACS images. Combined, this use of nearest-pixel flux densities provided at least one data point for \percentage{11} of the AT20G sources.  Nearest-pixel flux densities accounted for all three data points for only \percentage{6} of AT20G sources.

Nine AT20G sources were confused in MWACS to such an extent that it was impossible to reliably estimate their MWACS flux densities. These sources remained in the sample, but were not matched to any AT20G source; this is unlikely to significantly affect any of the conclusions we draw in this paper as they represent less than \percentage{1} of the final source count.

Henceforth, integrated or peak flux densities are referred to as `flux density measurements' to distinguish them from nearest-pixel flux densities which are referred to as `flux density estimates'.

The catalogue thus produced contained 1285 sources and was named `AT20G-MWACS'. This served as our sample for the successive studies.

\begin{table}
  \centering
  \caption{The number and proportion of AT20G sources for each possible combination of MWACS-3$\sigma$ frequency matches.}
  \label{tab:matchstats}
  \begin{tabular}{r l}
    \toprule
    Combination of MWACS-3$\sigma$        & Number of AT20G \\
    frequency matches                     & sources (\emph{\%}) \\
    \midrule
    \SI{180}{\mega\hertz}               &  874  (\emph{68.0}) \\
    \SIlist{150; 180}{\mega\hertz}       &  802  (\emph{62.4}) \\
    \SIlist{119; 150; 180}{\mega\hertz} &  693  (\emph{53.9}) \\
    Total                                 & 1285  (\emph{100})  \\
    \bottomrule
  \end{tabular}
\end{table}

\subsubsection{Reliability and completeness}
\label{Reliability and completeness}
Calculations of reliability and completeness were made using the results of the Monte Carlo simulations.

Reliability at each frequency band is given by the ratio of AT20G matches assumed real to total number of matches expected possible within the angular separation cutoff. For instance, at \SI{180}{\mega\hertz}, 874 AT20G sources had MWACS-3$\sigma$ matches within \SI{120}{\arcsec}, while 31 simulated sources had MWACS-3$\sigma$ matches within \SI{120}{\arcsec}. The reliability was thus \((874)/(874 + 31) = 0.966\) or \percentage{96.6}. Similarly the reliabilities at \SIlist{150; 119}{\mega\hertz} were 96.6 and \percentage{96.4}, respectively. Subsequently, the reliability of the entire catalogue could be considered the product of these factors: \percentage{90.0}.

All matches within the \SI{120}{\arcsec} cutoff at \SI{180}{\mega\hertz} were accepted, hence the completeness was \percentage{100} in this zone. All matches above \SI{120}{\arcsec} were excluded but the Monte Carlo simulations suggested that few or no genuine matches could occur beyond \SI{120}{\arcsec}, so the completeness would not be affected by the absence of matches in this zone.
Similar practices were followed at the remaining frequencies, hence the completeness of the entire catalogue can be considered close to \percentage{100}.

\subsection{Inclusion of additional flux density measurements and angular size information}

The AT20G catalogue contains flux density measurements at \SIlist{5; 8}{\giga\hertz} for 1138 of the 1285 sources in our sample. An existing extension to the catalogue \citep[AT20GHARC;][]{chhetri2013} provided flux densities at \SI{1.4}{\giga\hertz} and/or \SI{843}{\mega\hertz} for 1277 of our sources, with the \SI{1.4}{\giga\hertz} flux densities originating from the NRAO VLA Sky Survey \citep[NVSS;][]{condon1998} and the \SI{843}{\mega\hertz} flux densities from the Sydney University Molonglo Sky Survey \citep[SUMSS;][]{mauch2007} or the 2nd epoch Molonglo Galactic Plane Survey \citep[MGPS-2;][]{murphy2007}.

We further expanded the AT20G-MWACS catalogue to include additional flux densities from the MWA A priori Reference CatalOgue (MARCO)\footnote{
  Full details on how this catalogue was constructed, along with all of the scripts required for interested readers to generate the catalogue themselves are available at \href{http://cira.ivec.org/dokuwiki/doku.php/mwa/marco}{\texttt{http://cira.ivec.org/dokuwiki/doku.php/mwa/marco}}.
}. The MARCO was constructed by cross-matching NVSS and SUMSS with the following low frequency catalogues: the VLA Low-frequency Sky Survey catalogue \citep[VLSS;][]{cohen2007} at \SI{74}{\mega\hertz}, the Culgoora catalogue \citep{slee1995} at 80 and \SI{160}{\mega\hertz},
and the Molonglo Reference Catalogue \citep[MRC;][]{large1981} at \SI{408}{\mega\hertz}. Since the AT20G catalogue already contained NVSS and SUMSS identifications, it was trivial to match it with the MARCO. We expanded our AT20G-MWACS catalogue to include all MARCO flux density measurements, adding at least one additional flux density measurement for \percentage{72} of the AT20G sources.

\subsection{Harmonisation of catalogue uncertainties}

In order to use the uncertainties on the flux density measurements in various catalogues in a meaningful way when fitting, it was necessary to prevent flux densities with extremely small uncertainties overly influencing the final fit. In particular, the uncertainties on AT20G points were typically much smaller than the uncertainties on MWACS-3$\sigma$ points, which heavily biased the fitting procedures away from the low-frequency data. Consequently, for each of the three AT20G frequencies we added \percentage{10} of the integrated flux density in quadrature to the original uncertainty.

\section{Modelling spectral energy distributions}
\label{sec:modelling}

Since the dominant emission mechanism for radio sources is typically synchrotron radiation from relativistic electrons, many radio source SEDs can be accurately modelled by a power law
\begin{equation}
  S(\nu) = S_0 \left( \frac{\nu}{\nu_0} \right) ^ \alpha,
\end{equation}
or in log-log space,
\begin{equation}
  \log{S(\nu)} = \log{S_0} + \alpha \log \left( \frac{\nu}{\nu_0} \right).
\end{equation}
where $S$ is the flux density as a function of frequency $\nu$, $S_0$ is the flux density at the reference frequency, $\nu_0$, and $\alpha$ is the spectral index. For synchrotron processes $\alpha$ is typically around $-0.7$. We use $\nu_0 = \SI{1}{\giga\hertz}$ as the reference frequency throughout.

For many radio sources, this simple model breaks down due to frequency-dependent absorption (e.g., free-free absorption or synchrotron self-absorption) or a spectral break due to the source turning off.
Sources may also consist of several unresolved components, such as a flat-spectrum core dominating at higher frequencies, and steeper-spectrum jet-powered lobes dominating at low frequencies.

A straightforward and obvious modification to the model that can capture most of these behaviours is to extend it to a quadratic in log-log space and allow for a curvature parameter, $\phi$:

\begin{align}
  S(\nu) &= S_0 \left( \frac{\nu}{\nu_0} \right) ^ \alpha 10 ^ {\phi \log ^ 2 \left( \frac{\nu}{\nu_0} \right)}, \\
  \log{S(\nu)} &= \log{S_0} + \alpha \log \left( \frac{\nu}{\nu_0} \right) + \phi \log ^ 2 \left( \frac{\nu}{\nu_0} \right) .
\end{align}

The curvature, $\phi$, has a value of zero or near-zero for a linear power law SED. It takes larger absolute values the more sharply the spectral energy distribution curves in the log space domain. If positive, the curve is opening upward (convex) and a trough may be seen within the data range; if negative, the curve is opening downward (concave) and a peak may be seen within the data range. The linear term, $\alpha$, simply represents the slope of the curve at the reference frequency, $\nu_0$. For negative curvature, a more positive $\alpha$ implies a peak at high frequencies and a negative $\alpha$ implies a peak at low frequencies (and vice versa for a trough).

It is also useful to know the frequency of the turning point (peak or, more rarely, trough), $\nu_{\mathrm{turn}}$. This is simply
\begin{equation}
 \nu_{\mathrm{turn}} = 10^{ -\frac{\alpha}{2\phi} }. \label{eq:quadroots}
\end{equation}

\subsection{Implementation}

\subsubsection{Fitting the models}

Fitting with robust uncertainties was done via a Monte Carlo scheme, whereby 500 realisations of each SED were generated by perturbing each flux density $S$ by $\updelta S$. For flux density \emph{measurements}, $\updelta S$ was drawn from a normal distribution with mean equal to zero and standard deviation equal to the uncertainty on $S$. For flux density \emph{estimates}, $\updelta S$ was drawn from a normal distribution with mean equal to the nearest-pixel flux density and standard deviation equal to the local RMS.

In each realisation both the linear and quadratic models as described above were fit using the \function{polyfit()} function from the \package{NumPy} Python package, and the parameters describing those fits were recorded.

After the simulations were complete, each parameter for each fitted model had an ensemble of 500 possible values. The central value for each parameter was assumed to be the median of the values found by simulation, and the uncertainty to be the semi-interhexile range. The use of median and semi-interhexile range were preferred because these terms approximated the mean and standard deviation for a Gaussian distribution and were more resistant to the effects of any outliers present.

The goodness of fit parameter, $\chi^2$, was calculated for each realisation of each model. Flux density estimates were not used in this process, i.e., the flux density estimates were used to guide the fitting process and arrive at values for the parameters, but the resultant fits were judged only on the basis of how they met the flux density measurements. This prevented fits from being punished when the uncertainties on the flux density estimates were large, as often happened otherwise.

$\sub{\nu}{turn}$ was calculated for each realisation, however the ensemble of values was only used to calculate the uncertainty on $\sub{\nu}{turn}$, the value itself being determined from $\alpha$ and $\phi$ using Equation~\ref{eq:quadroots}.

\subsubsection{Selection of superior model}
\label{sec:selection-model}

Model selection was achieved using one of three possible criteria, checked in the following order: significance of curvature, then value of reduced chi-squared ($\subsup{\chi}{red}{2}$) \emph{or} value of the corrected Akaike information criterion (AICc).
If the first test deemed the modelled curvature was significant, the second test acted to determine which model was a better fit.
A breakdown of the sample subsets by selected model and test responsible is provided in Table~\ref{tab:selection-reasons}.

\begin{table}
  \centering
  \caption{The subsets of our 1285 sources identified as either linear or quadratic, explicitly listing the variable which was specifically tested when the linear model was selected as superior.}
  \label{tab:selection-reasons}
  \begin{tabular}{r r l c c c}
    \toprule
    ~    & Total & (\emph{\%})   & \(\phi\) & \(\chi^{2}_\mathrm{red}\) & AICc \\
    \midrule
    Linear    &  457 & (\emph{35.6}) & 353 & 41 & 63 \\
    Quadratic &  828 & (\emph{64.4}) & -   & -  & -   \\
    \bottomrule
  \end{tabular}

\end{table}

If the curvature of the quadratic fit was not deemed significant, the linear fit was immediately accepted with no further checks. Quadratic curvature was considered insignificant if the \(\phi\) parameter was less than its uncertainty, i.e. $\frac{|\phi|}{\Delta\phi}<1$, this representing a quadratic fit consistent with no curvature from the Monte Carlo simulations. As a natural consequence of this definition, our final sample had no sources with \(0 < |\phi| < 0.02\). This check was responsible for \percentage{77.2} of the linear model selections.

When the curvature was considered significant, one of the two following checks was applied depending on the number of data points available.
For four or fewer data points, the AICc could not be determined for both fits and instead for each fit we found the value of chi-squared per degree of freedom
\begin{equation}
  \subsup{\chi}{red}{2} = \frac{\chi^2}{\operatorname{dof}} = \frac{\chi^2}{N - k}
\end{equation}
where $N$ is the number of data points and $k$ is the number of fitted parameters (either 2 or 3 for the linear or quadratic model, respectively). If there were zero degrees of freedom available \(\subsup{\chi}{red}{2}\) was marked invalid. If both fits were valid and had \(\subsup{\chi}{red}{2}\) within $\pm0.05$ of 1, the linear model was chosen, otherwise the fit with \(\subsup{\chi}{red}{2}\) closest to 1 was chosen.
If only one fit was valid, that model was automatically chosen.
If neither fit was valid, the linear model was chosen.
This check was responsible for \percentage{9.0} of the linear fit selections.

For 5 or more data points, it was possible to calculate the value of AICc for both fits by
\begin{equation}
  \operatorname{AICc} = \chi^{2} + 2k + \frac{2k(k + 1)}{N - k - 1}
\end{equation}
where $N$ and $k$ are defined as above. This check was responsible for \percentage{13.8} of the linear fit selections.

The (uncorrected) Akaike information criterion is given by \(\operatorname{AIC} = \chi^{2} + 2k\) when the variance on each point fitted is the same for all models tested.
The AICc is a corrected form of the AIC which introduces a weighting factor to account for fits with $N \sim k^2$ \citep{Burnham2002}, such as in our case where $N$ would range from \numrange{5}{12}, and is valid for univariate linear models such as ours. The fit with the lower value of AICc was selected as superior.

It was important to use the AICc technique when possible as it accounted for the number of degrees of freedom in a more thorough manner than $\subsup{\chi}{red}{2}$ alone, and generally speaking AIC is preferred for comparing models when a true model may not be among the candidates. Since the quadratic model had no physical motivation, this was particularly applicable in our case.

\subsection{Assessing fit quality}

The chosen fit was classified as `good' or `poor' based on assessment of two metrics which were empirically determined to be indicative of a fit that still appeared ill-suited to the data.

\begin{enumerate}
  \item If the \(\subsup{\chi}{red}{2}\) of the chosen fit was greater than 50 the source was flagged as poor. This check was used even for sources selected by the AICc technique. If there had been too few data points to calculate \(\subsup{\chi}{red}{2}\), in which case the source had been marked invalid and the linear model chosen, the source was also flagged as poor.
  \item If the turning point of the spectrum, $\sub{\nu}{turn}$, had a relative uncertainty greater than 0.5 and fell within our observed frequency domain the source was flagged as poor.
\end{enumerate}

It was possible for a source to be flagged by both of these checks. Table~\ref{tab:fitQual} lists a breakdown of the quality subsets.

\begin{table}
  \centering
  \caption{The subsets of our 1285 sources with each fit quality flag. Every source is flagged either `good' or `poor'; at least one of the two tests, on \(\chi^{2}_\mathrm{red}\) and \(\nu_{\textrm{turn}}\), was failed by a poor source.}
  \label{tab:fitQual}
  \begin{tabular}{r r l c c}
    \toprule
    ~    & Total & (\emph{\%})   & Bad \(\chi^{2}_\mathrm{red}\) & Bad \(\nu_{\textrm{turn}}\) \\
    \midrule
    Good &  994 & (\emph{77.7}) & -   & -   \\
    Poor &  291 & (\emph{22.3}) & 183 & 131 \\
    \bottomrule
  \end{tabular}

\end{table}

It is important to acknowledge that a source may have been poorly fit by both the linear and quadratic models, even if it had very precise flux density measurements, simply because the SED could not be simplified to either of our models. The study of such sources is beyond the scope of
this work and we consider them no further.

\section{Results: Advantages of the \texorpdfstring{\mbox{\large \(\balpha\)}-\mbox{\large \(\bphi\)}}{alpha-phi} diagram}
\label{sec:alpha-phi-intro}

\begin{figure}
  \centering
  \includegraphics[width = 0.95\columnwidth]{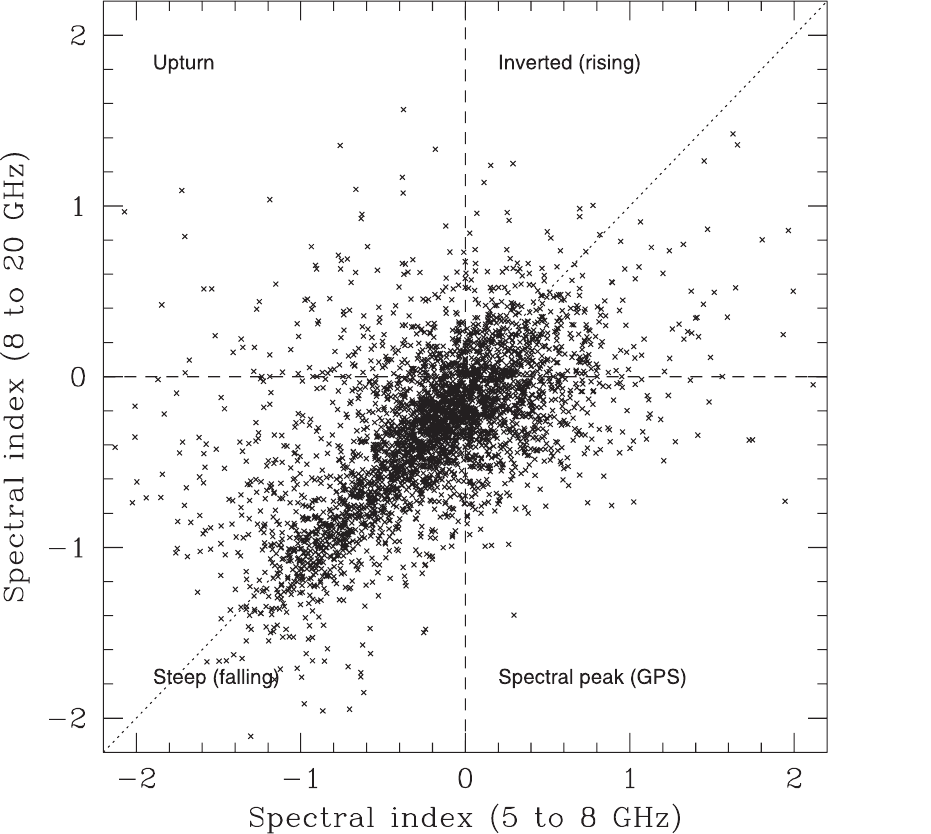}
  \caption{Radio colour-colour (or \emph{two-colour}) diagram from AT20G \citep*[fig. 15;][]{AT20Gsource}. Crosses show the \roughly{3800} sources observed at all three frequencies. The dotted line indicates the loci of sources with a pure power-law SED.}
  \label{fig:AT20G-RCC}
\end{figure}

Once the parameters $\alpha$ and $\phi$ have been determined for each source, with robust uncertainties on each, we can consider how to use these parameters to classify radio sources. As a starting point we consider the radio colour-colour (RCC) plot, developed by \cite{kesteven1977} though adapted from \cite{bolton1969}. As an example, the RCC diagram by \cite{AT20Gsource} for AT20G sources is reproduced in Fig.~\ref{fig:AT20G-RCC}. It takes the spectral indices ($\alpha_1$ and $\alpha_2$) as observed through two `windows' of frequency and plots them against each other, essentially mapping the source in a two-dimensional space based on the gradients it exhibits between flux densities at three or four observed frequencies, assuming that the spectral index is constant across each window.

The quadrant of the RCC diagram into which a source falls can be used to qualify its broad-spectrum behaviour, classifying it as steep, upturned, inverted, peaked, or flat. Sources which follow a linear power law relationship will fall along the line $\alpha_1 = \alpha_2$.

When the spectral information on a source is limited to three or four spectral points within an order of magnitude of each other, the RCC diagram succinctly captures the known information. However as the spectra of the sources are mapped in greater detail and over a greater range of wavelengths, the RCC plot often fails to capture important features. For example, a source which appears peaked does so because in one window it exhibits a positive spectral index and in the other a negative. Unless the turning point just happens to fall precisely between the two spectral windows, however, the true character of the spectral shape is masked.

Across the two orders of magnitude of frequency in which our observations were placed, we expected to see an even wider variety of spectral behaviour than that observed over the three frequencies used in the AT20G survey.
While fairly sophisticated modelling would be possible with the rich dataset we had synthesised, our aim was to find a two-dimensional summary of the spectral properties of our source population which none the less utilised all spectral information.

\begin{figure}
  \centering
  \includegraphics[width = \columnwidth]{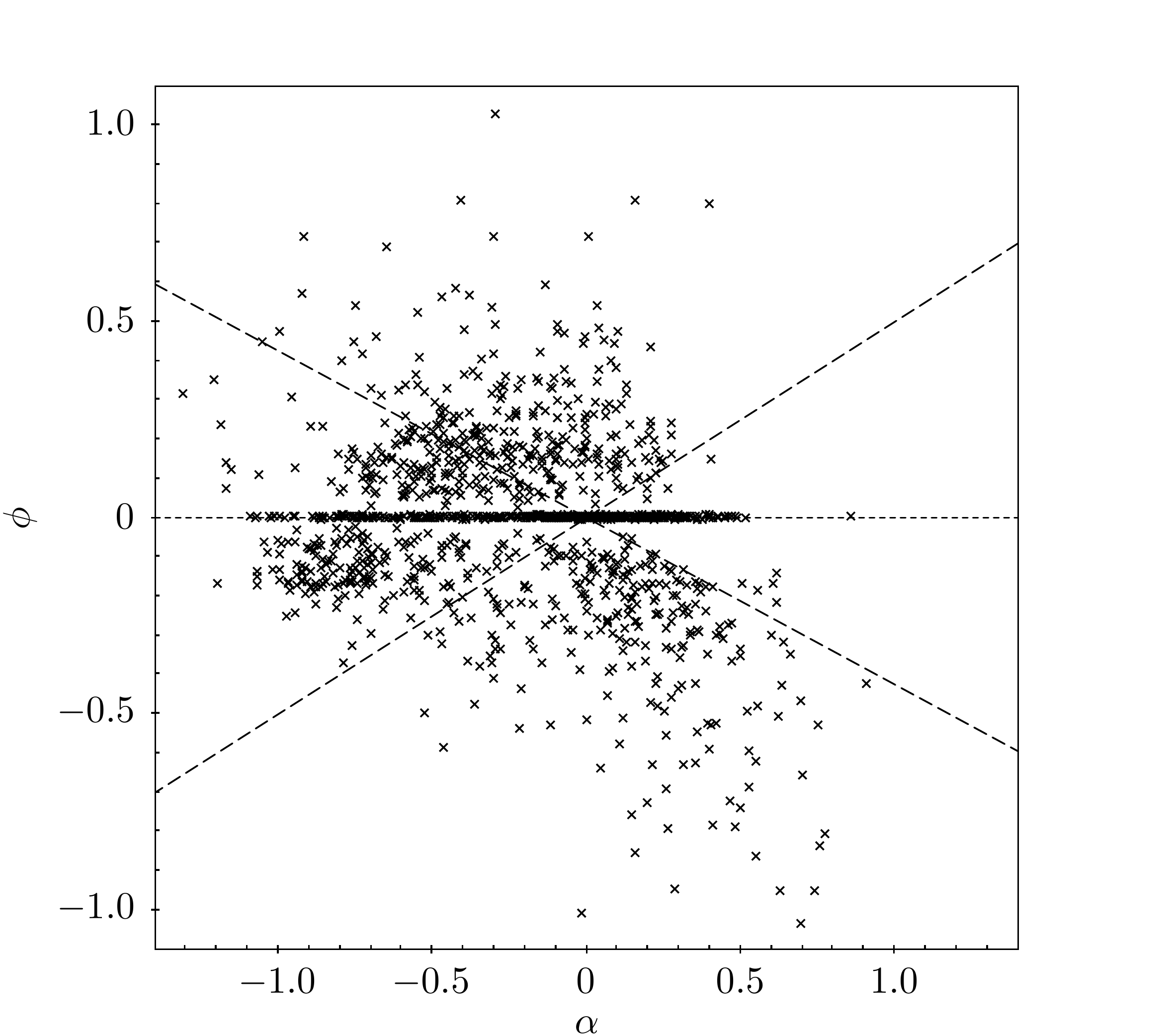}
  \caption{Simple \(\alpha\)-\(\phi\) diagram for all 1285 AT20G-MWACS sources. The dotted horizontal line is comparable to the dotted diagonal line in Fig.~\ref{fig:AT20G-RCC}, i.e. with a power-law SED. Sources with \(\phi = 0\) have been slightly scattered off this axis for visibility. The dashed diagonal lines separate the curved subclasses (see text), and are loosely comparable to the dashed quadrant lines in Fig.~\ref{fig:AT20G-RCC}.}
  \label{fig:alpha-phi}
\end{figure}

As the name would suggest, an \(\alpha\)-\(\phi\) diagram plots the fitted parameters \(\alpha\) and \(\phi\) against one another for each source, which is already far more intuitive than using the same parameter, \(\alpha\), as viewed through two neighbouring windows and mapping it into two dimensions. The \(\alpha\)-\(\phi\) diagram for our sample of sources is shown in Fig.~\ref{fig:alpha-phi}, though as it features sources with both good and poor fit qualities, care must be taken in its interpretation.

Upon initial inspection of the \(\alpha\)-\(\phi\) diagram, it is clear that there are three distinct groups of sources. In the upper half, positive \(\phi\) indicates the quadratic is tending towards convex behaviour, implying what is referred to as an upturned source. Along a line through the middle, a \(\phi\) of zero marks those sources which display little to no curvature, and can be modelled by a single spectral index (easily read from the \(\alpha\) axis). Finally, in the lower half, negative \(\phi\) indicates concave behaviour, implying a peaked source.

It may appear that the \(\alpha\)-\(\phi\) diagram is little more than an RCC diagram rotated \SI{45}{\degree} clockwise. While the RCC diagram attempts to map points based on their spectral shape, it does so in a very forced way. Sources supposedly well-fit by a linear power law lie along the diagonal, but it is unclear how far from the diagonal this population extends. Furthermore, the populations of upturned or peaked sources are taken from quadrants of the diagram under the assumption the two windows can well-describe the curvature.
Consider the \(\alpha\)-\(\phi\) diagram, where with the value of \(\phi\) we have quantified curvature into a single variable, and what the RCC diagram achieved with a diagonal, the \(\alpha\)-\(\phi\) diagram does with a much more readable horizontal. It is easier and more intuitive to pick populations out of the \(\alpha\)-\(\phi\) diagram since it is now clear precisely which sources can be considered linear, and this draws a boundary.

Conveniently, both \(\alpha\) and \(\phi\) typically range from about \numrange{-1}{+1}, allowing intuitively comfortable scales on the diagrams.

\subsection{Further classification of sources on the \texorpdfstring{\(\balpha\)-\(\bphi\)}{alpha-phi} diagram}

Four to five classes have historically been used in conjunction with the RCC diagram: one for each quadrant (steep, upturned, inverted, peaked), and in some cases a fifth near the origin (flat). In order to facilitate easier understanding of the representative meaning for the location of a source on the \(\alpha\)-\(\phi\) diagram, we sought to adapt these existing class names as much as possible. To that end we propose a system of three superclasses and three subclasses, only requiring the introduction of one new identifier. In total, this leads to nine unique radio spectral classes (RSCs).

\begin{table}
  \centering
  \caption{The nine radio spectral classes (RSC). Each describes the rough shape of the optimum fitted SED model of a source. Observe how the steep RSCs tend downwards, the inverted tend upwards, the upturned have convex curvature, and the peaked have concave curvature. In shorthand, the RSCs are: Us, Uf, Ui, Ls, Lf, Li, Ps, Pf, and Pi.}
  \label{tab:special-instructions}
  \begin{tabular}{c *{3}{>{\centering\arraybackslash}m{0.2\columnwidth}}}
    \toprule
    ~ & s & f & i \\
    \midrule
    U & \includegraphics[width=0.2\columnwidth]{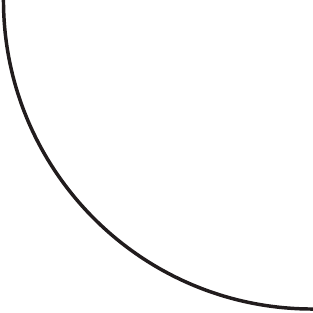} & \includegraphics[width=0.2\columnwidth]{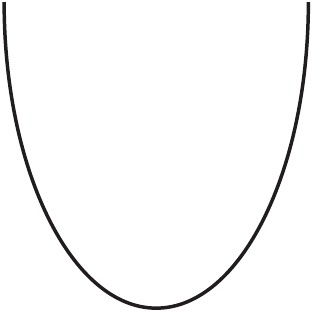} & \includegraphics[width=0.2\columnwidth]{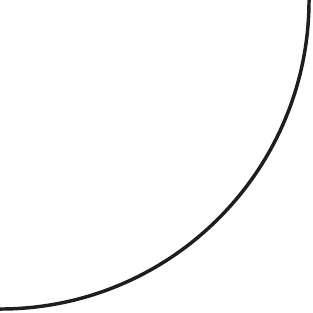} \\
    L & \includegraphics[width=0.2\columnwidth]{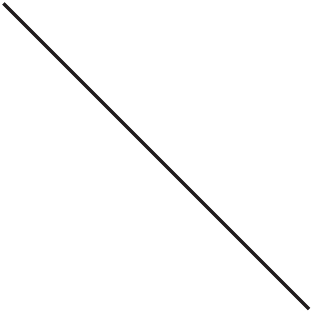} & \includegraphics[width=0.2\columnwidth]{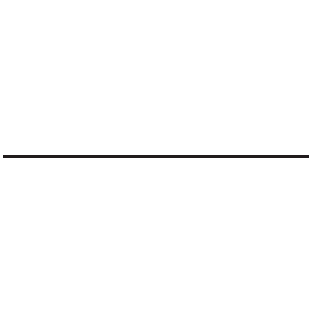} & \includegraphics[width=0.2\columnwidth]{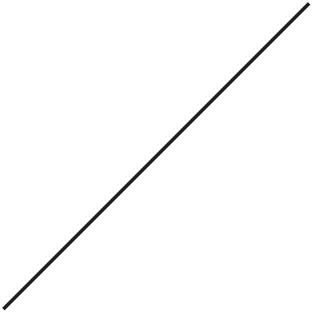} \\
    P & \includegraphics[width=0.2\columnwidth]{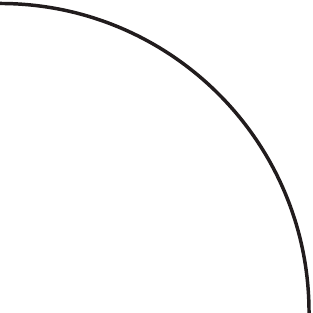} & \includegraphics[width=0.2\columnwidth]{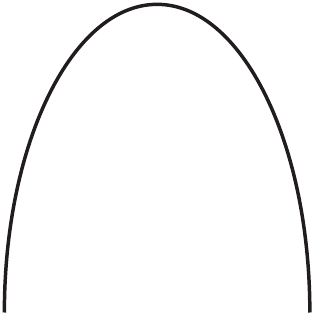} & \includegraphics[width=0.2\columnwidth]{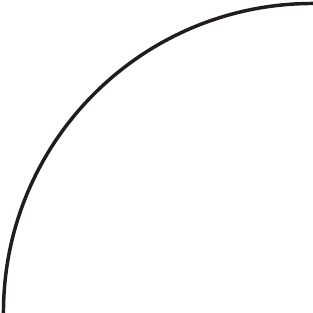} \\
    \bottomrule
  \end{tabular}
\end{table}

As visualised in Table~\ref{tab:special-instructions}, the superclasses are upturned (U), linear (L), and peaked (P). The subclasses are steep (s), flat (f), and inverted (i). By combining each superclass with each subclass, we form the nine RSCs. For convenience, we may refer to the upturned and peaked superclasses jointly as `curved'.

Superclasses are relatively well defined, as summarised in Table~\ref{tab:define-sup}. As mentioned earlier, \(\phi\) very conveniently represents the direction of curvature of a source, so when positive the source is upturned and when negative the source is peaked. The only point of ambiguity is where one chooses to recognize that \(\phi\) is close enough to zero as to be essentially modelling a simple power law. This varies slightly with \(\alpha\), as sources with large \(\alpha\) can have a large \(\phi\) and still be better fit by a linear model whereas a source with a smaller \(\alpha\) but the same  \(\phi\) would be better fit by the quadratic. As stated in Section~\ref{sec:selection-model}, we allowed the statistical methods to decide when a curved fit was superior to a linear one, aiming to minimize the presence of empirically-determined cutoff points.

\begin{table}
  \centering
  \caption{Summary of the boundaries between each of the nine possible RSCs.}
  \label{tab:define-classes}
  \begin{subtable}[b]{\columnwidth}
    \centering
     \caption{}
    \label{tab:define-sup}
    \begin{tabular}{c c}
      \toprule
      Superclass & Definition         \\
      \midrule
      U          & \(\phi > 0\)       \\
      L          & \(\phi \approx 0\) \\
      P          & \(\phi < 0 \)      \\
      \bottomrule
    \end{tabular}
  \end{subtable}%

  ~%

  \begin{subtable}[b]{\columnwidth}
    \centering
    \caption{}
    \label{tab:define-sub}
    \begin{tabular}{r c}
      \toprule
      Subclass & Definition (Freq. in \si{\giga\hertz}) \\
      \midrule
      U \(\begin{cases} \text{s} \\ \text{f} \\ \text{i} \end{cases}\) & \(\begin{aligned} &\phantom{\nu_{\mathrm{turn}}<\;} 15 < \nu_{\mathrm{turn}} \phantom{.} \\ 0.1 \leq \;& \nu_{\mathrm{turn}} \leq 15 \\ \nu_{\mathrm{turn}} < 0.1 \phantom{\;<\;} &\phantom{\nu_{\mathrm{turn}}} \end{aligned}\) \\
      \midrule
      L \(\begin{cases} \text{s} \\ \text{f} \\ \text{i} \end{cases}\) &  \(\begin{aligned} \alpha \leq -0.5 \phantom{\;<\;} &\phantom{\alpha} \\ -0.5 < \;& \alpha \leq 0 \\ &\phantom{\alpha>\;} 0 < \alpha \phantom{-.5} \end{aligned}\) \\
      \midrule
      P \(\begin{cases} \text{s} \\ \text{f} \\ \text{i} \end{cases}\) & \(\begin{aligned} \nu_{\mathrm{turn}} < 0.1 \phantom{\;<\;} &\phantom{\nu_{\mathrm{turn}}} \\ 0.1 \leq \;& \nu_{\mathrm{turn}} \leq 15 \\ &\phantom{\nu_{\mathrm{turn}}<\;} 15 < \nu_{\mathrm{turn}} \phantom{.} \end{aligned}\) \\
      \bottomrule
    \end{tabular}
  \end{subtable}
\end{table}

Subclasses are somewhat less rigorously separable, and in fact have slightly different definitions between the linear and curved superclasses, summarised in Table~\ref{tab:define-sub}. As such, the source must be assigned a superclass before the correct subclass can be determined (hence the distinction). In keeping with the established conventions for RCC interpretation, we consider sources with no curvature to be linear-steep (Ls) when \(\alpha \leq -0.5\), linear-flat (Lf) when \(-0.5 < \alpha \leq 0\), and linear-inverted (Li) when \(\alpha > 0\).

When it comes to the curved superclasses, the steep and inverted subclasses are indicative of sources which are displaying curvature in keeping with the superclass, but are still generally tending in a steep or inverted manner. For sources in these four classes -- upturned-steep (Us), upturned-inverted (Ui), peaked-steep (Ps), and peaked-inverted (Pi) -- we do not actually see a clear sign of a trough or a peak, simply that the source has statistically-significant curvature. The flat subclass of curved superclasses indicates that the turning point of the parabola -- the trough or peak -- lies within the domain of catalogued frequencies. These two classes -- upturned-flat (Uf) and peaked-flat (Pf) -- are, in essence, defining the \emph{truly} upturned and peaked spectra, where we can see evidence of this entire energy distribution shape. Since we wished to generally have flux densities available to confirm the Uf and Pf classes, they were restricted to sources for which \(\sub{\nu}{turn}\) fell between \SI{100}{\mega\hertz} and \SI{15}{\giga\hertz}.

As we calculated \(\sub{\nu}{turn}\) regardless of whether the spectrum was reaching a turning point within our range, we could also use it as an extrapolated turning point frequency. A source with an extrapolated trough above \SI{15}{\giga\hertz} or a peak below \SI{100}{\mega\hertz} was considered to be in the subclass steep, and a trough below \SI{100}{\mega\hertz} or a peak above \SI{15}{\giga\hertz} inverted. These two limits, together with Equation~\ref{eq:quadroots}, define the diagonal lines on Fig.~\ref{fig:alpha-phi}. Since the limits are not equidistant from the reference frequency of \SI{1}{\giga\hertz}, and furthermore are dependent on the range of frequencies observed, the diagonals are not perpendicular to one another.

It should be noted that depending on the domain of catalogued frequencies, the distinction between subclasses is far more variable than the distinction between superclasses. Indeed, the RSC system requires a fairly wide frequency domain in order to properly characterise the source. As we will show in Section~\ref{sec:discussion}, however, we consider this an extremely necessary consideration -- even with the RCC diagram, we can rarely be accurate in our assumptions about spectral forms without a broad data set.

\section{The catalogue}
\label{sec:catalogue}

\subsection{Selection bias}

The AT20G-MWACS catalogue is an extension of the original AT20G catalogue, thus it only contains sources with a flux density above \SI{40}{\milli\jansky\per\beam} at \SI{20}{\giga\hertz}, this being the detection limit of the AT20G survey \citep{AT20Gsource}. The detection limit of MWACS-3$\sigma$ is approximately \SI{50}{\milli\jansky\per\beam} at \SI{180}{\mega\hertz}. Therefore, sources with inverted spectra which are already weak at \SI{20}{\giga\hertz} have gone completely undetected in the MWACS-3$\sigma$ source-finding. For example, no source with a true linear spectrum that has a spectral index of $\alpha = 0.5$ and a \SI{20}{\giga\hertz} flux density less than roughly \SI{520}{\milli\jansky\per\beam} will have been directly detected by MWACS-3$\sigma$. Such AT20G sources are still in our final catalogue, with quoted flux density estimates, but this is of limited use.

\subsection{RSC populations}

\begin{figure}
  \centering
  \includegraphics[width = \columnwidth]{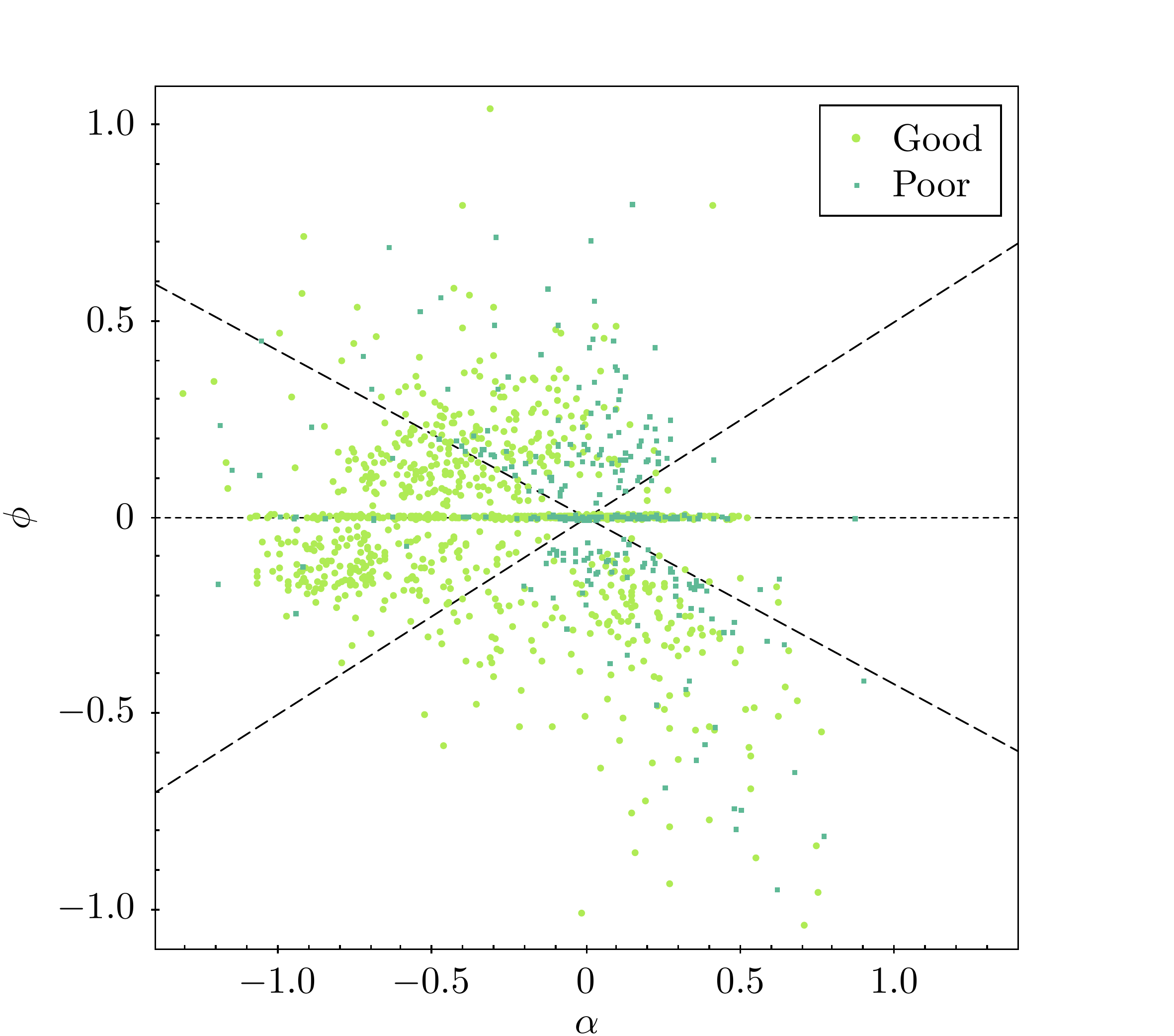}
  \caption{Same as Fig.~\ref{fig:alpha-phi}, except sources now discriminated by fit quality. Lime green circles: Sources with good model fits. Turquoise squares: Sources with poor model fits. Note the clustering of poor fits in the linear superclass; these were typically sources which fell back on a linear model due to unusual SEDs or too few measurements to calculate \(\subsup{\chi}{red}{2}\).}
  \label{fig:alpha-phi-fitQual}
\end{figure}

The \(\alpha\)-\(\phi\) diagram for our sample as it appears in Fig.~\ref{fig:alpha-phi} is not yet useful for population studies as it continues to feature sources with poor fit quality. Fig.~\ref{fig:alpha-phi-fitQual} is a reproduction of Fig.~\ref{fig:alpha-phi}, illustrating the spread of the poor fit quality sources.
The majority of these poor fit quality sources lie in the region of positive $\alpha$ because they are only detected at high frequencies and are poorly constrained by flux density estimates taken from MWACS, thus introducing significant uncertainties on the fits for both models.

\begin{figure}
  \centering
  \includegraphics[width = \columnwidth]{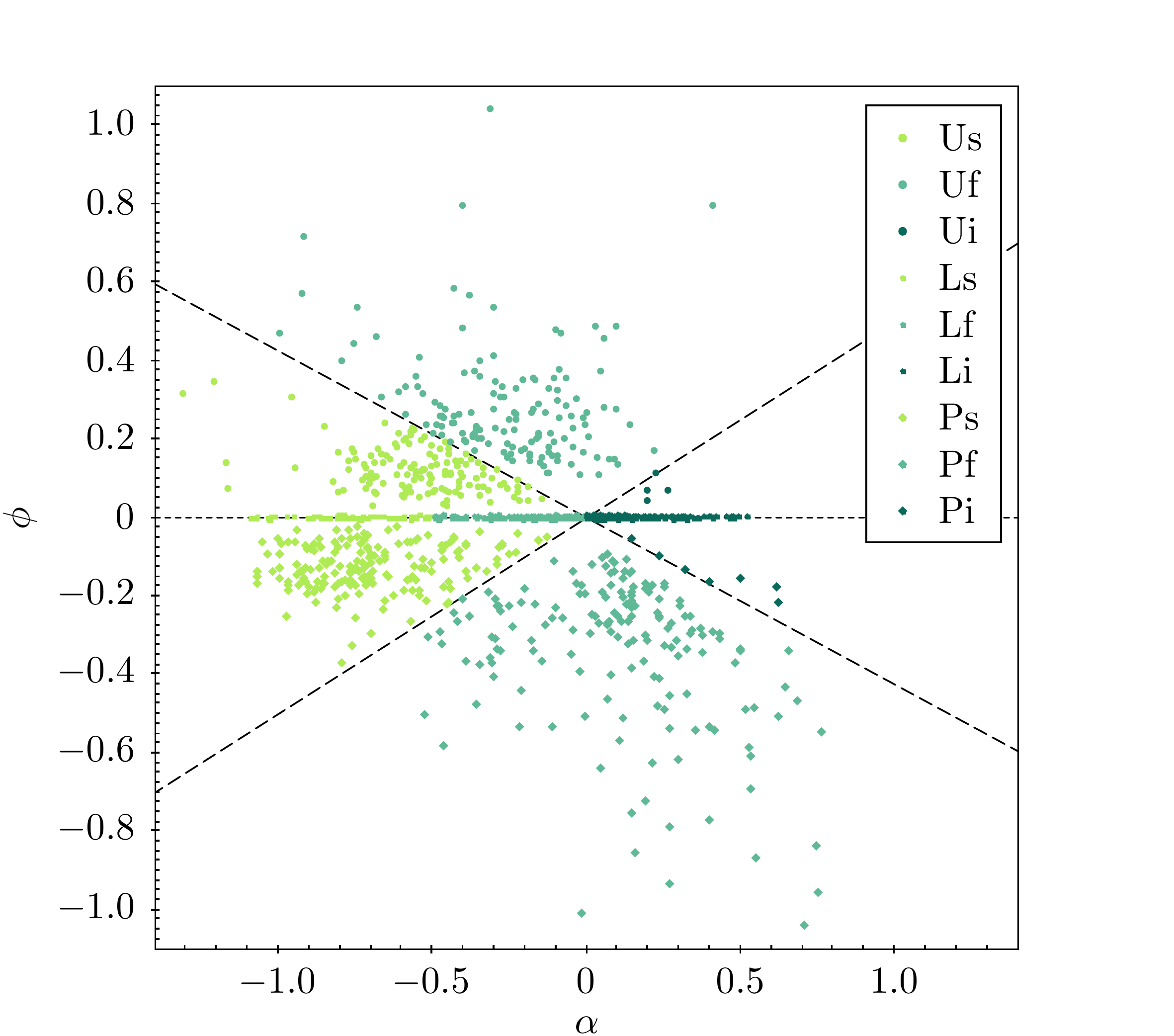}
  \caption{Similar to Fig.~\ref{fig:alpha-phi}; an \(\alpha\)-\(\phi\) diagram for the 994 AT20G sources with good fit qualities, differentiating superclasses by shape and subclasses by colour. Lime green: Steep. Turquoise: Flat. Dark green: Inverted. Circles: Upturned. Squares: Linear. Diamonds: Peaked.}
  \label{fig:alpha-phi-fitQual_g}
\end{figure}

If we exclude the sources with poor fit quality we are left with Fig.~\ref{fig:alpha-phi-fitQual_g}, and can begin to study relative population sizes for these sources. These values are provided in Table~\ref{tab:populations}.

\begin{table}
  \centering
  \caption{Absolute sizes of various RSC subsets, with percent of the total sample of 994 good fit quality sources provided in parentheses.}
  \label{tab:populations}
  \begin{tabular}{c *{3}{>{\centering\arraybackslash}m{0.2\columnwidth}}}
    \toprule
    ~ & s & f & i \\
    \midrule
    U & \multicolumn{2}{c}{\(\leftarrow\) 275~(\emph{27.7}) \(\rightarrow\)} & 4~(\emph{0.4}) \\
    L & 80~(\emph{8.05}) & \multicolumn{2}{c}{\(\leftarrow\) 283~(\emph{28.5}) \(\rightarrow\)}\\
    P & \multicolumn{3}{c}{\(\longleftarrow\) 352~(\emph{35.4}) \(\longrightarrow\)} \\
    \bottomrule
  \end{tabular}
\end{table}

\subsubsection{Peaked-steep, -flat, and -inverted}

Constituting just over a third of our sample, the three classes in the lower half of the \(\alpha\)-\(\phi\) diagram all display either a peak within our frequency domain or a tendency towards a peak outside of our frequency domain (see Table~\ref{tab:special-instructions}). By spectral shape alone, these would be compact steep-spectrum (CSS) and gigahertz peaked-spectrum (GPS) sources \citep{ODea:98}. Presently believed to be the early stages of evolution for radio galaxies, these sources begin with a high-frequency peak that evolves over time to lower frequencies, moving through the Pf then Ps classes. The reason a peak is observed is not fully understood, and while synchrotron self-absorption is believed to play a role, free-free absorption is also possible in some sources \citep{Callingham:15}. While at high frequencies we can see the steep negative spectral shape created by the lobes, at lower frequencies this emission starts being absorbed by the galaxy itself, and creates a sudden drop in the observed flux density. The larger the galaxy, the lower the peak frequency.

\subsubsection{Linear-steep}

This sizeable class contains galaxies dominated by synchrotron emission from their jets and lobes. The emission from the core of the galaxy is far weaker than that from the lobes, thus the flux density becomes quite low at higher frequencies. In these galaxies we can usually see both jets, as they are directed at right-angles to our line of sight.

\subsubsection{Upturned-steep and -flat}

Towards these classes are galaxies which show signs of activity in the galactic nucleus at high frequencies in conjunction with the aforementioned lobe activity at low frequencies. The inverted emission at high frequencies is likely to attain maximum somewhere above \SI{20}{\giga\hertz}. Some proportion of Uf sources may be restarted radio galaxies, with a set of highly extended lobes from long-prior activity as well as a newer set of lobes closer to the renewed-active core. Unfortunately, identifying these sources typically requires a high degree of visual inspection, seeking cases where multiple low-frequency components surround a compact high-frequency core. In some cases the outer set of lobes is merged with the inner set due to the poor resolution of the low-frequency survey, and the flux density is overestimated for these sources, creating an upturned shape.

\subsubsection{Linear-flat and -inverted}

In some cases the sources in the Lf class have frequency points so scattered that a horizontal linear model is statistically the better fit, but this does not account for the majority of these sources once the poor fits are excluded. In the case of these galaxies, it is possible the jet could be pointing towards our line of sight \citep[i.e. these sources are blazars;][]{Schmidt:68}, thus appearing significantly brighter due to relativistic beaming.

\subsubsection{Upturned-inverted}

Virtually unpopulated within our catalogue, possibly because no sources with this spectral shape physically exist, though more likely due to the selection biases in our sampling technique. Fitting to these sources would require MWACS-3$\sigma$ sources much weaker than AT20G, thus placing them in the realm of using flux density estimates rather than flux density measurements. The fluctuations in flux density estimates are large, as are the uncertainties, so the fitting functions would be unlikely to settle on a model which is flat in this domain.

\subsection{Catalogue columns}

The complete AT20G-MWACS catalogue is available online, along with an extended version that includes the statistics of our model fits and classification results. Column numbers, names, units, and descriptions for the extended catalogue are provided in Table~\ref{tab:cat-columns}.

\begin{table*}
  \centering
  \caption{Column information for the AT20G-MWACS catalogue. If applicable, each column description notes which catalogue the original data was drawn from. A value of 0.0 in the S180M, S150M, or S120M columns indicates that calculations were performed using the flux density estimates from the L180M, L150M, or L120M columns respectively.}
  \label{tab:cat-columns}
  \begin{tabular}{c c c l}
    \toprule
    Number & Name & Units & Description \\
    \midrule
1  & AT20G           & JHHMMSS+DDMMSS              & AT20G -- IAU designation \\
2  & RAJ2000         & \si{\degree}                & AT20G -- Right Ascension J2000 \\
3  & DEJ2000         & \si{\degree}                & AT20G -- Declination J2000 \\
4  & S20G            & \si{\milli\jansky\per\beam} & AT20G -- Peak brightness at \SI{20}{\giga\hertz} \\
5  & e\_S20G         & \si{\milli\jansky\per\beam} & Uncertainty on S20G with an additional \SI{10}{\percent} of S20G added in quadrature \\
6  & e\_S20G\_orig   & \si{\milli\jansky\per\beam} & AT20G -- Uncertainty on S20G \\
7  & S8G             & \si{\milli\jansky\per\beam} & AT20G -- Peak brightness at \SI{8.6}{\giga\hertz} \\
8  & e\_S8G          & \si{\milli\jansky\per\beam} & Uncertainty on S8G with an additional \SI{10}{\percent} of S8G added in quadrature \\
9  & e\_S8G\_orig    & \si{\milli\jansky\per\beam} & AT20G -- Uncertainty on S8G \\
10 & S5G             & \si{\milli\jansky\per\beam} & AT20G -- Peak brightness at \SI{4.8}{\giga\hertz} \\
11 & e\_S5G          & \si{\milli\jansky\per\beam} & Uncertainty on S5G with an additional \SI{10}{\percent} of S5G added in quadrature \\
12 & e\_S5G\_orig    & \si{\milli\jansky\per\beam} & AT20G -- Uncertainty on S5G \\
13 & S1G             & \si{\milli\jansky}          & NVSS -- Integrated \SI{1.4}{\giga\hertz} flux density \\
14 & e\_S1G          & \si{\milli\jansky}          & NVSS -- Mean uncertainty on S1G \\
15 & S843M           & \si{\milli\jansky}          & SUMSS or MGPS-2 -- Integrated \SI{843}{\mega\hertz} flux density \\
16 & e\_S843M        & \si{\milli\jansky}          & SUMSS or MGPS-2 -- Uncertainty on S843M \\
17 & S408M           & \si{\milli\jansky}          & MRC -- Integrated flux density at \SI{408}{\mega\hertz} \\
18 & e\_S408M        & \si{\milli\jansky}          & MRC -- Uncertainty on S408M \\
19 & S180M           & \si{\milli\jansky}          & MWACS-3$\sigma$ -- Integrated island flux at \SI{180.48}{\mega\hertz} \\
20 & e\_S180M        & \si{\milli\jansky}          & MWACS-3$\sigma$ -- Uncertainty on S180M \\
21 & S150M           & \si{\milli\jansky}          & MWACS-3$\sigma$ -- Integrated island flux at \SI{149.76}{\mega\hertz} \\
22 & e\_S150M        & \si{\milli\jansky}          & MWACS-3$\sigma$ -- Uncertainty on S150M \\
23 & S120M           & \si{\milli\jansky}          & MWACS-3$\sigma$ -- Integrated island flux at \SI{119.04}{\mega\hertz} \\
24 & e\_S120M        & \si{\milli\jansky}          & MWACS-3$\sigma$ -- Uncertainty on S120M \\
25 & S160M           & \si{\milli\jansky}          & CCA -- Integrated flux density at \SI{160}{\mega\hertz} \\
26 & e\_S160M        & \si{\milli\jansky}          & CCA -- Uncertainty on S160M \\
27 & S80M            & \si{\milli\jansky}          & CCA -- Integrated flux density at \SI{80}{\mega\hertz} \\
28 & e\_S80M         & \si{\milli\jansky}          & CCA -- Uncertainty on S80M \\
29 & S74M            & \si{\milli\jansky}          & VLSS -- Integrated flux density at \SI{74}{\mega\hertz} \\
30 & e\_S74M         & \si{\milli\jansky}          & VLSS -- Uncertainty on S74M \\
31 & L180M           & \si{\milli\jansky}          & MWACS-3$\sigma$ -- Value of nearest pixel in MWACS image \\
32 & e\_L180M        & \si{\milli\jansky}          & MWACS-3$\sigma$ -- Local RMS in MWACS image \\
33 & L150M           & \si{\milli\jansky}          & MWACS-3$\sigma$ -- Value of nearest pixel in MWACS image \\
34 & e\_L150M        & \si{\milli\jansky}          & MWACS-3$\sigma$ -- Local RMS in MWACS image \\
35 & L120M           & \si{\milli\jansky}          & MWACS-3$\sigma$ -- Value of nearest pixel in MWACS image \\
36 & e\_L120M        & \si{\milli\jansky}          & MWACS-3$\sigma$ -- Local RMS in MWACS image \\
37 & alpha1\_20      &                             & Spectral index between \SIlist{1; 20}{\giga\hertz} \\
38 & alpha5\_8       &                             & Spectral index between \SIlist{4.8; 8.6}{\giga\hertz} \\
39 & alpha8\_20      &                             & Spectral index between \SIlist{8.6; 20}{\giga\hertz} \\
40 & alpha180M\_1G   &                             & Spectral index between \SI{180}{\mega\hertz} and \SI{1}{\giga\hertz} \\
41 & freq\_turn      & \si{\giga\hertz}            & Frequency of the turning point of the quadratic $\nu_\text{turn}$ \\
42 & e\_freq\_turn   & \si{\giga\hertz}            & Uncertainty on the frequency of the turning point of the quadratic $\nu_\text{turn}$ \\
43 & fluxRefFit      & \si{\giga\hertz}            & $S_0$ (flux at \SI{1}{\giga\hertz}) of the best-fit model \\
44 & e\_fluxRefFit   & \si{\giga\hertz}            & Uncertainty on $S_0$ (flux at \SI{1}{\giga\hertz}) of the best-fit model \\
45 & alphaFit        &                             & $\alpha$ of the best-fit model \\
46 & e\_alphaFit     &                             & Uncertainty on $\alpha$ of the best-fit model \\
47 & phiFit          &                             & $\phi$ of the best-fit model \\
48 & e\_phiFit       &                             & Uncertainty on $\phi$ of the best-fit model \\
49 & chiSqFit        &                             & $\chi^2$ of the best-fit model \\
50 & e\_chiSqFit     &                             & Uncertainty on $\chi^2$ of the best-fit model \\
51 & redChiSqFit     &                             & $\chi^2_\text{red}$ of the best-fit model \\
52 & e\_redChiSqFit  &                             & Uncertainty on $\chi^2_\text{red}$ of the best-fit model \\
53 & AICcLin         &                             & AICc of the linear model \\
54 & AICcQuad        &                             & AICc of the quadratic model \\
55 & numData         &                             & Number of measured data points available to the fits \\
56 & fitClass        & [ULP][sfi]                  & Radio spectral class (RSC) flag assigned to the source (see text) \\
57 & fitClassInt     &                             & Integer representation of RSC: from $-4$ to $+4$ represent Ps through to Ui \\
58 & fitQual         & [gp][x-][t-]                & Fit quality flag assigned to the source (see text) \\
59 & fitQualInt      &                             & Integer representation of fit quality: 0 (g{-}{-}), 1 (px-), 2 (p-t), 3 (pxt) \\
    \bottomrule
  \end{tabular}
\end{table*}

\section{Discussion}
\label{sec:discussion}

\subsection{Comparison to high frequency}

\subsubsection{Positive curvature}

\begin{figure}
  \centering
  \begin{minipage}[b]{\columnwidth}
    \includegraphics[width=\columnwidth]{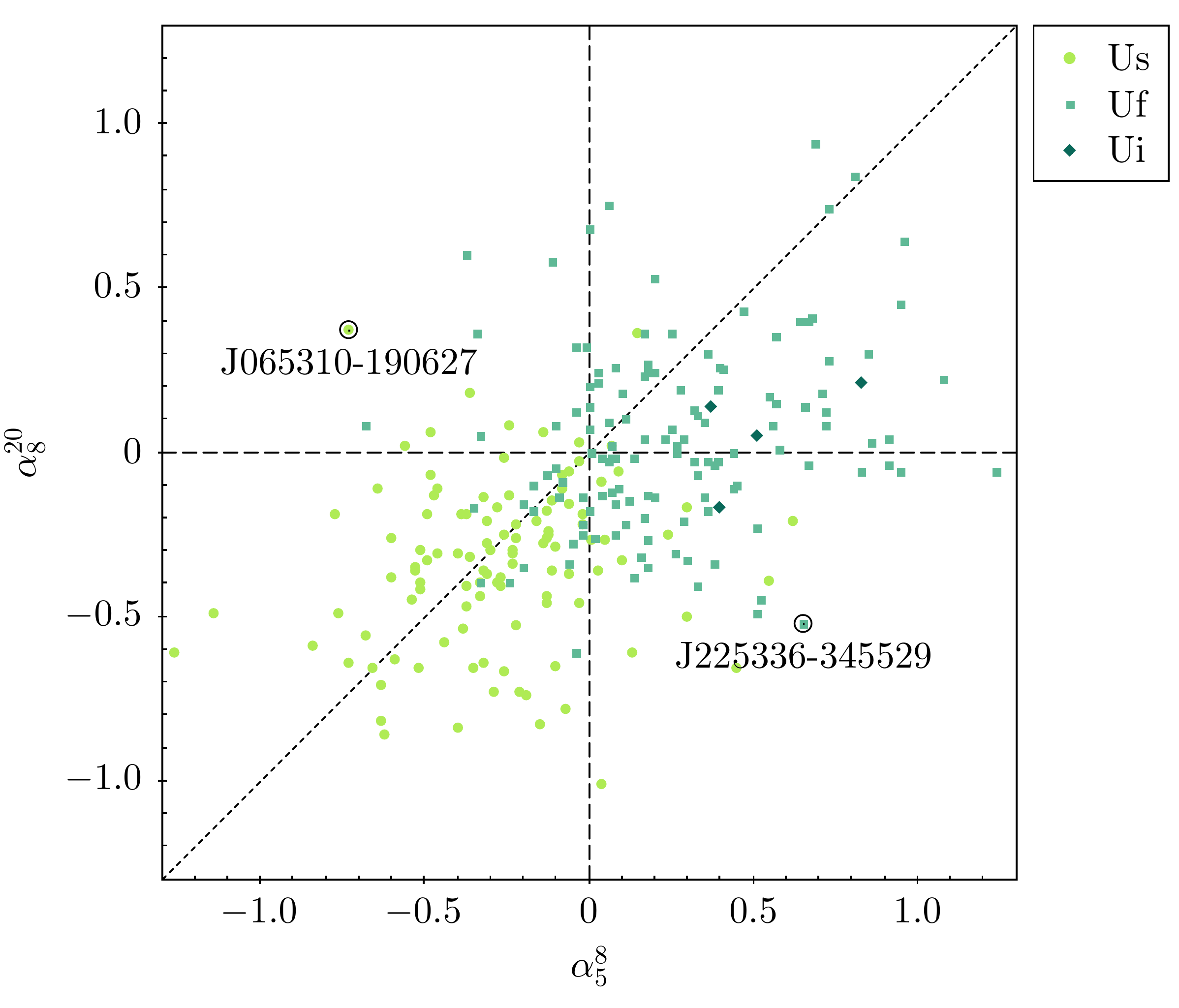}
    \footnotesize (a) RCC diagram from \SIrange{5}{8}{\giga\hertz} and \SIrange{8}{20}{\giga\hertz}, plotting the upturned sources from our good fit quality sample. Lime green circles: Us. Turquoise squares: Uf. Dark green diamonds: Ui. Four sources are beyond axis ranges.
  \end{minipage}

  \begin{minipage}[b]{0.41\columnwidth}
    \centering
    \includegraphics[width=\columnwidth]{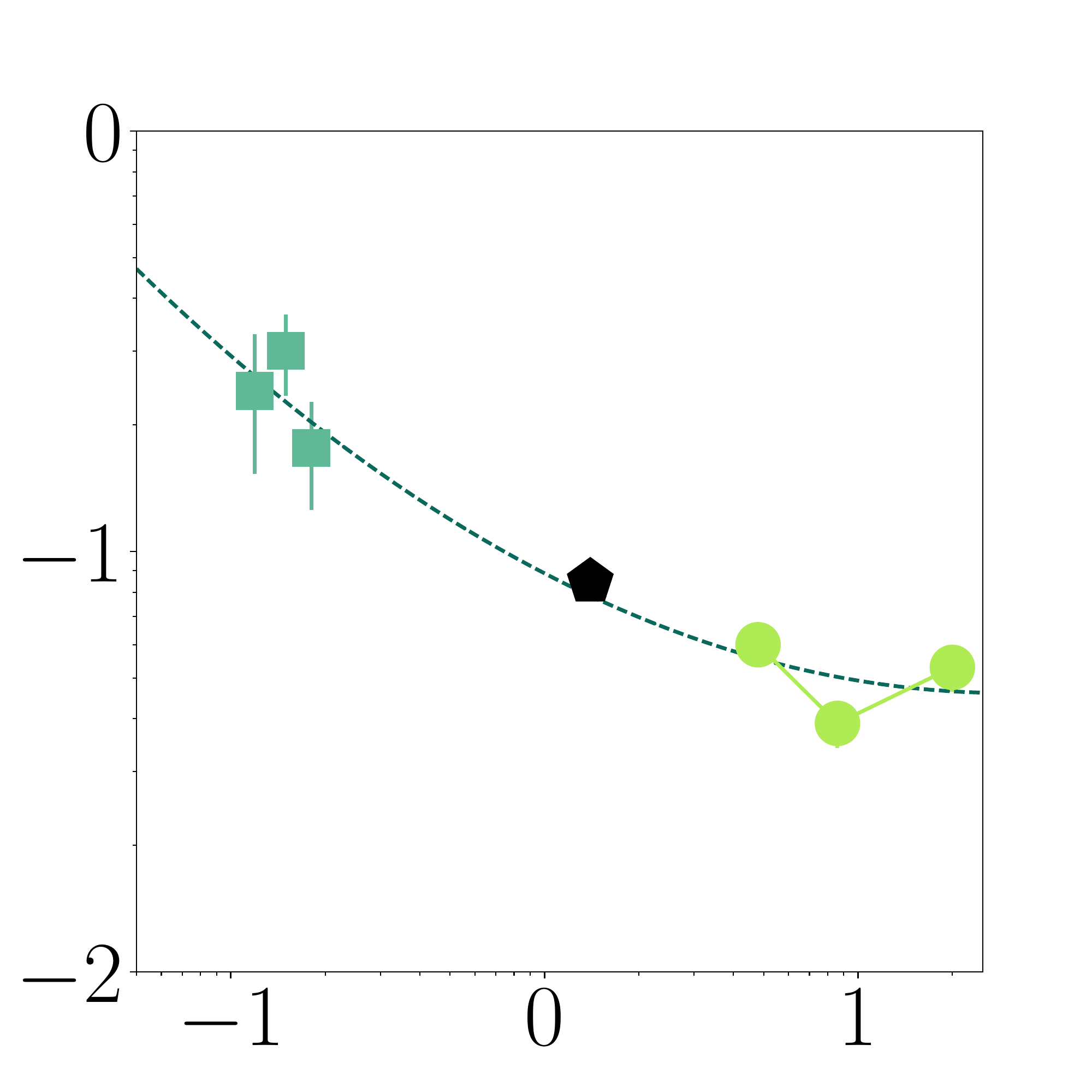}
    \footnotesize (b) J065310-190627
  \end{minipage}%
  \begin{minipage}[b]{0.41\columnwidth}
    \centering
    \includegraphics[width=\columnwidth]{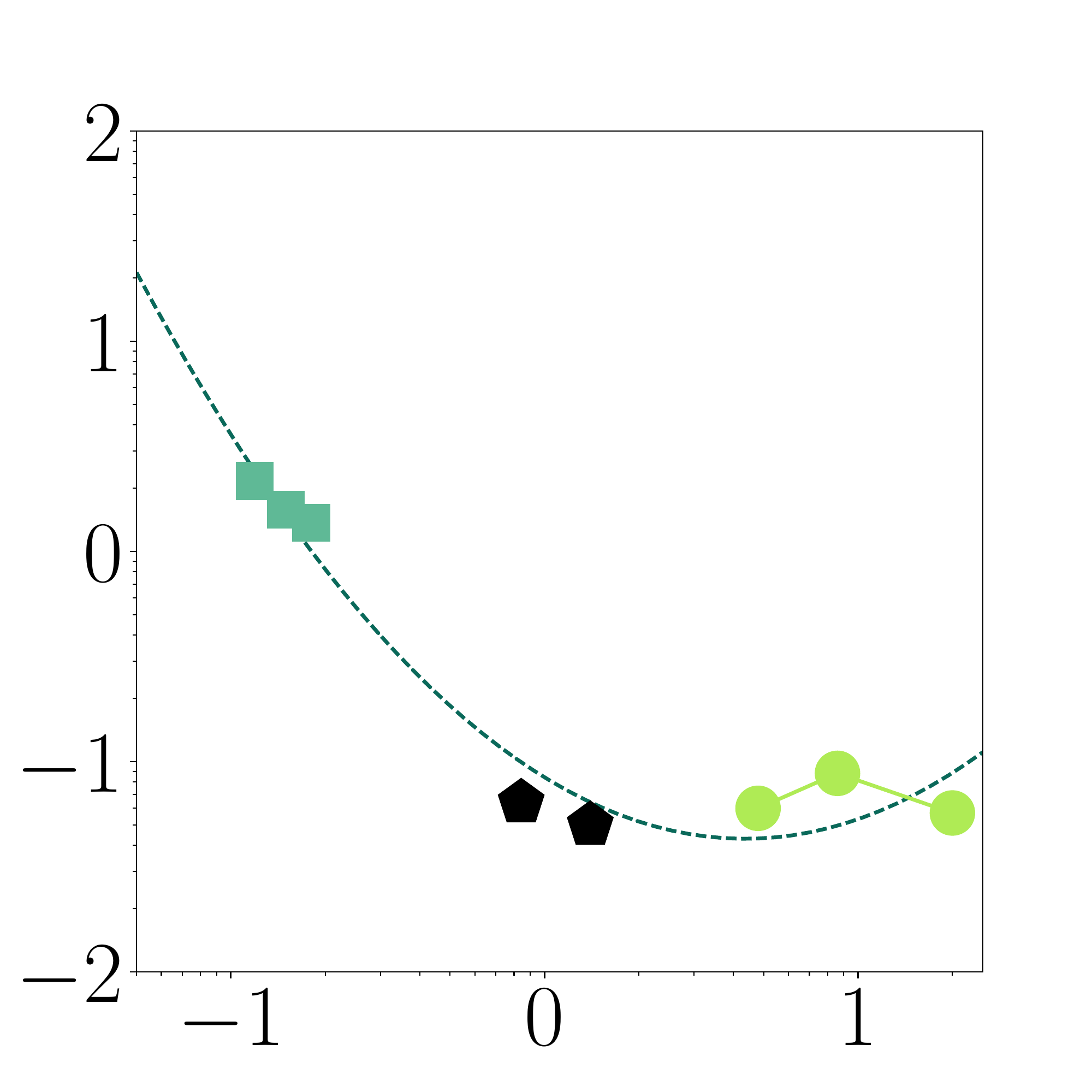}
    \footnotesize (c) J225336-345529
  \end{minipage}%

  \caption{RCC diagram and two example SEDs identified by the AT20G source name. For these SEDs the $x$-axis has units $\log \left( \nu/\nu_0 \right)$ and the $y$-axis has units $\log \left( S/\SI{1}{\jansky} \right)$ -- Lime green circles joined by solid line: AT20G measurements. Turquoise squares: MWACS measurements. Black pentagons: Other measurements. Dashed dark green line: Chosen fitted model.}
  \label{fig:alpha5_8-alpha8_20-fitClass_U}
\end{figure}

Fig.~\ref{fig:alpha5_8-alpha8_20-fitClass_U}a takes all sources with positive curvature from Fig.~\ref{fig:alpha-phi-fitQual_g} and plots their location by AT20G spectral index data and style and colour by our RSCs. As these are the sources with positive curvature we would expect them to all lie above the diagonal on the RCC diagram if the three AT20G measurements had alone been enough to properly characterise the SED shape. The scatter of points in the RCC diagram is extremely wide; the quadrants do not correlate well with our new \(\alpha\)-\(\phi\) classification, and it is apparent that at high frequencies these sources appear to have a wide variety of spectral shapes.

Two-thirds of the Uf sources are below the diagonal, and indeed a significant number are in the lower-right quadrant which \cite{AT20Gsource} would call `peaked' in the AT20G survey. This demonstrates that an RCC diagram for this subset of sources totally misclassifies most of them when using high-frequency data alone.

Fig.~\ref{fig:alpha5_8-alpha8_20-fitClass_U}b and~\ref{fig:alpha5_8-alpha8_20-fitClass_U}c show how sources can become misclassified by variations in the AT20G flux density measurements (much larger than the uncertainty on these points).
In Fig.~\ref{fig:alpha5_8-alpha8_20-fitClass_U}b the AT20G data alone places the source in the upturned quadrant (note the solid lime green line taking the shape of a trough), but the broad-frequency spectrum reveals this SED shape was misleading and the turnover frequency is likely outside of our domain.
In Fig.~\ref{fig:alpha5_8-alpha8_20-fitClass_U}c the AT20G data alone places the source in the peaked quadrant, but the broad-frequency spectrum reveals the source in fact may be displaying a trough in our domain, entirely contradictory to the high-frequency conclusion. We note however that the lower frequency AT20G is non-contemporaneous to the \SI{20}{\giga\hertz} flux density measurements and there may some intrinsic flux variations if these observations are probing the cores of galaxies.

\begin{figure}
  \centering
  \begin{minipage}[b]{\columnwidth}
    \includegraphics[width=\columnwidth]{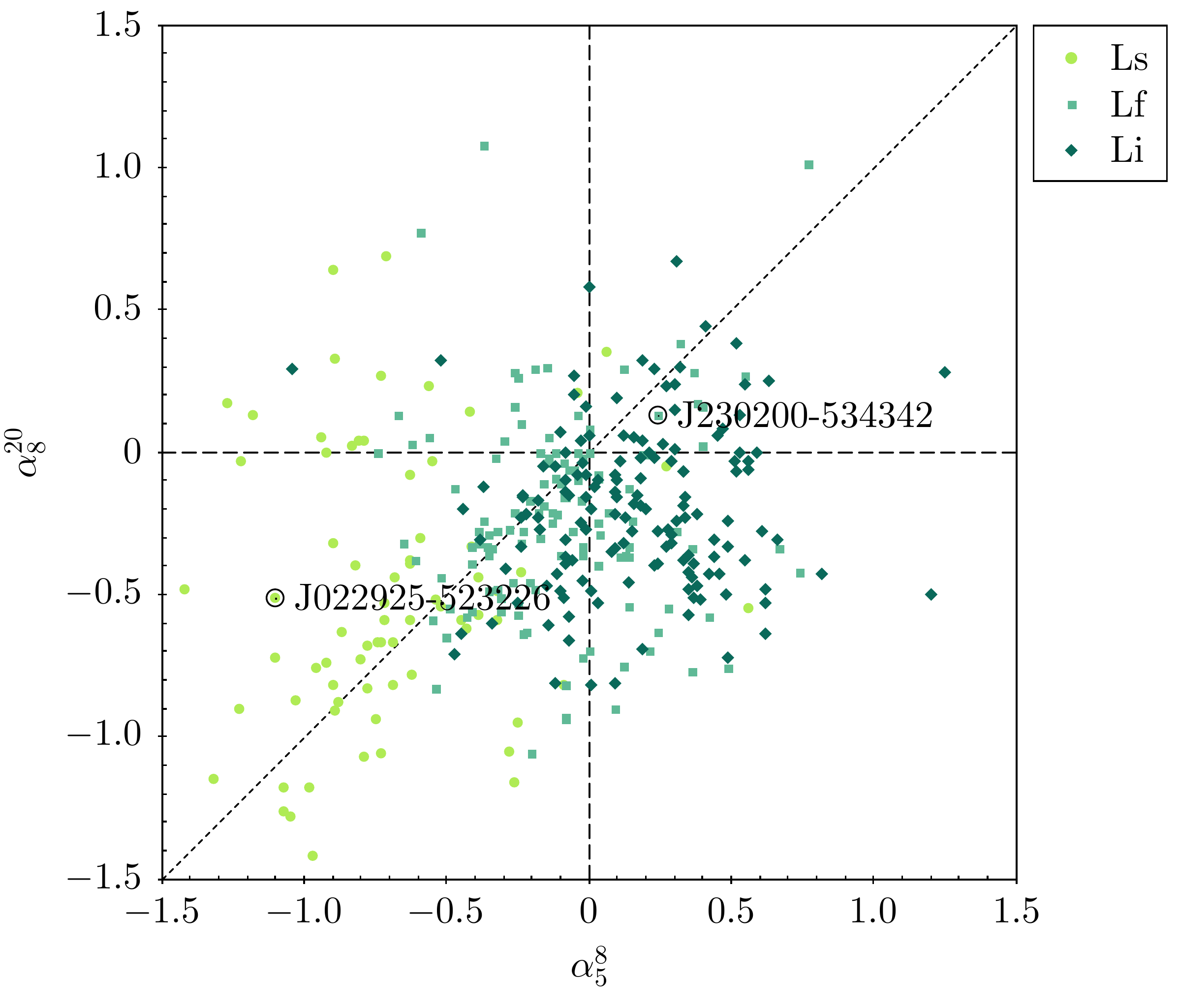}
    \footnotesize (a) Similar to Fig.~\ref{fig:alpha5_8-alpha8_20-fitClass_U}a, plotting the linear sources from our good fit quality sample. Lime green circles: Ls. Turquoise squares: Lf. Dark green diamonds: Li. Four sources are beyond axis ranges.
  \end{minipage}

  \begin{minipage}[b]{0.41\columnwidth}
    \centering
    \includegraphics[width=\columnwidth]{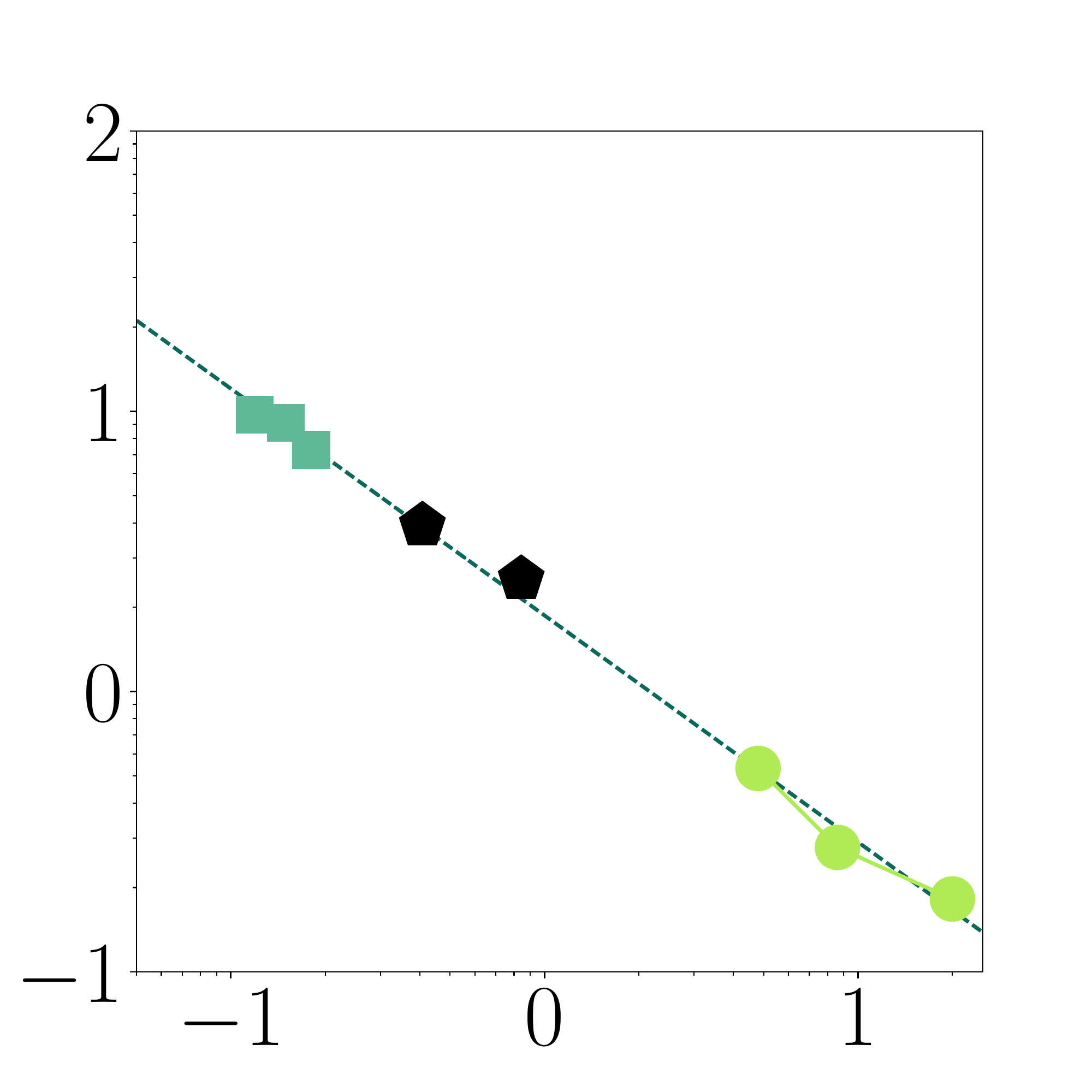}
    \footnotesize (b) J022925-523226
  \end{minipage}%
  \begin{minipage}[b]{0.41\columnwidth}
    \centering
    \includegraphics[width=\columnwidth]{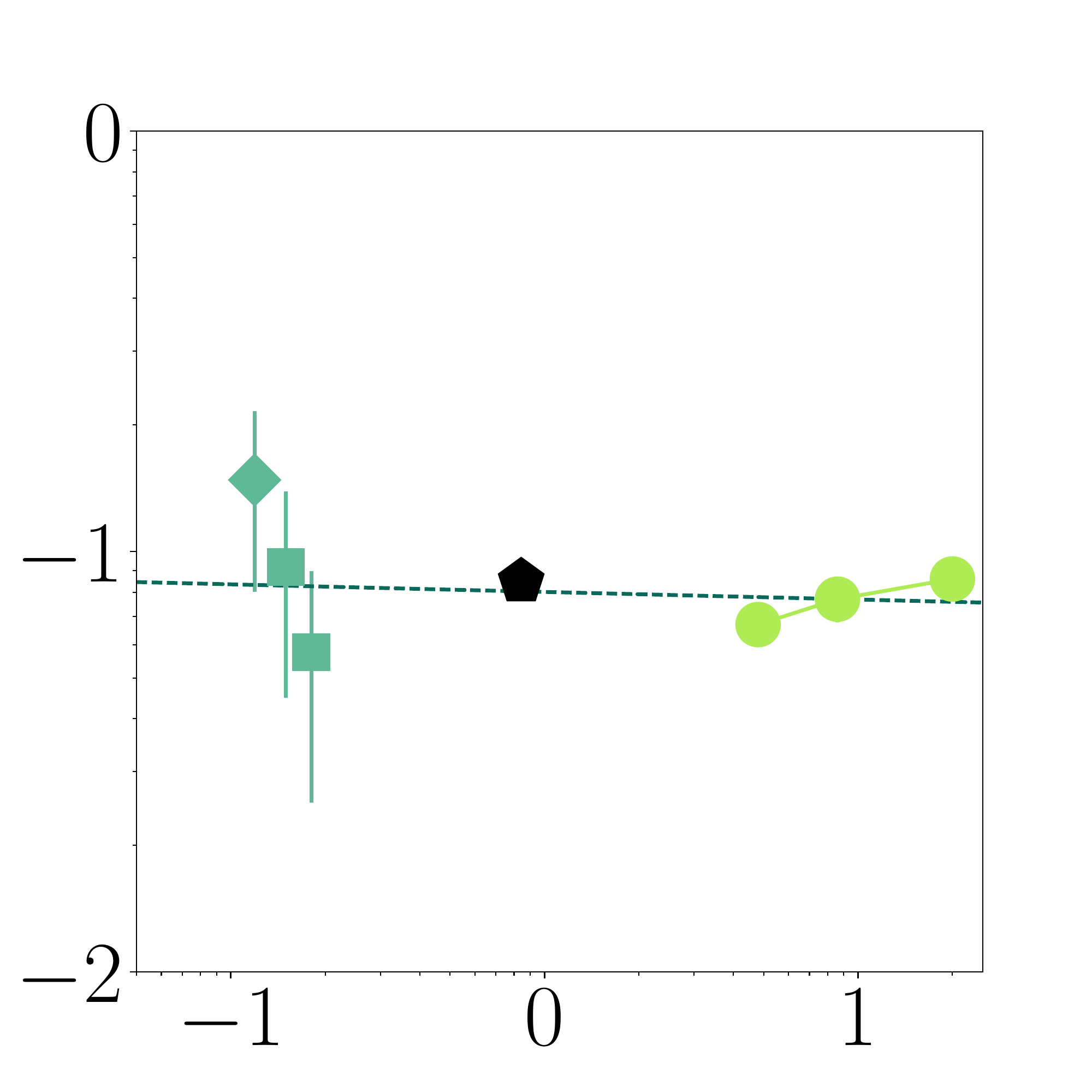}
    \footnotesize (c) J230200-534342
  \end{minipage}%

  \caption{Similar to Fig.~\ref{fig:alpha5_8-alpha8_20-fitClass_U}, with addition for these SEDs -- Turquoise diamonds: MWACS estimates.}
  \label{fig:alpha5_8-alpha8_20-fitClass_L}
\end{figure}

\subsubsection{No curvature}

Fig.~\ref{fig:alpha5_8-alpha8_20-fitClass_L} takes all sources with zero curvature from Fig.~\ref{fig:alpha-phi-fitQual_g} and plots their location by AT20G spectral index data and style and colour by our RSCs. As they have no curvature, we would expect them to lie along the diagonal in the RCC diagram if the AT20G frequencies alone were sufficient to characterise the sources. Again, the high scatter is obvious, however it is also clear that many Ls sources fall in the steep quadrant, as we might have expected. In total, \percentage{51.6} of these linear sources fall in the quadrant we would expect to see them (steep for Ls and Lf, inverted for Li). This indicates that the RCC diagram is particularly effective when only plotting linear sources, although this is rarely going to be the case in practice.

The linear model fitted to the source in Fig.~\ref{fig:alpha5_8-alpha8_20-fitClass_L}b has \(\spindex{fit} = -0.81 \pm 0.02\), which is significantly different to the two spectral indices calculated from AT20G of \(\spindex[8]{5} = -1.1\) and \(\spindex[20]{8} = -0.51\). Only the broad-frequency fit has the resiliency to go almost unaffected by the unevenness of the high-frequency data points.

Fig.~\ref{fig:alpha5_8-alpha8_20-fitClass_L}c shows how, even in a worst case where the fit can only provide a very general idea of the spectral behaviour due to the complexity of the spectrum, the AT20G classification of the behaviour can still be entirely contradictory to the broad-spectrum view.

\begin{figure}
  \centering
  \begin{minipage}[b]{\columnwidth}
    \includegraphics[width=\columnwidth]{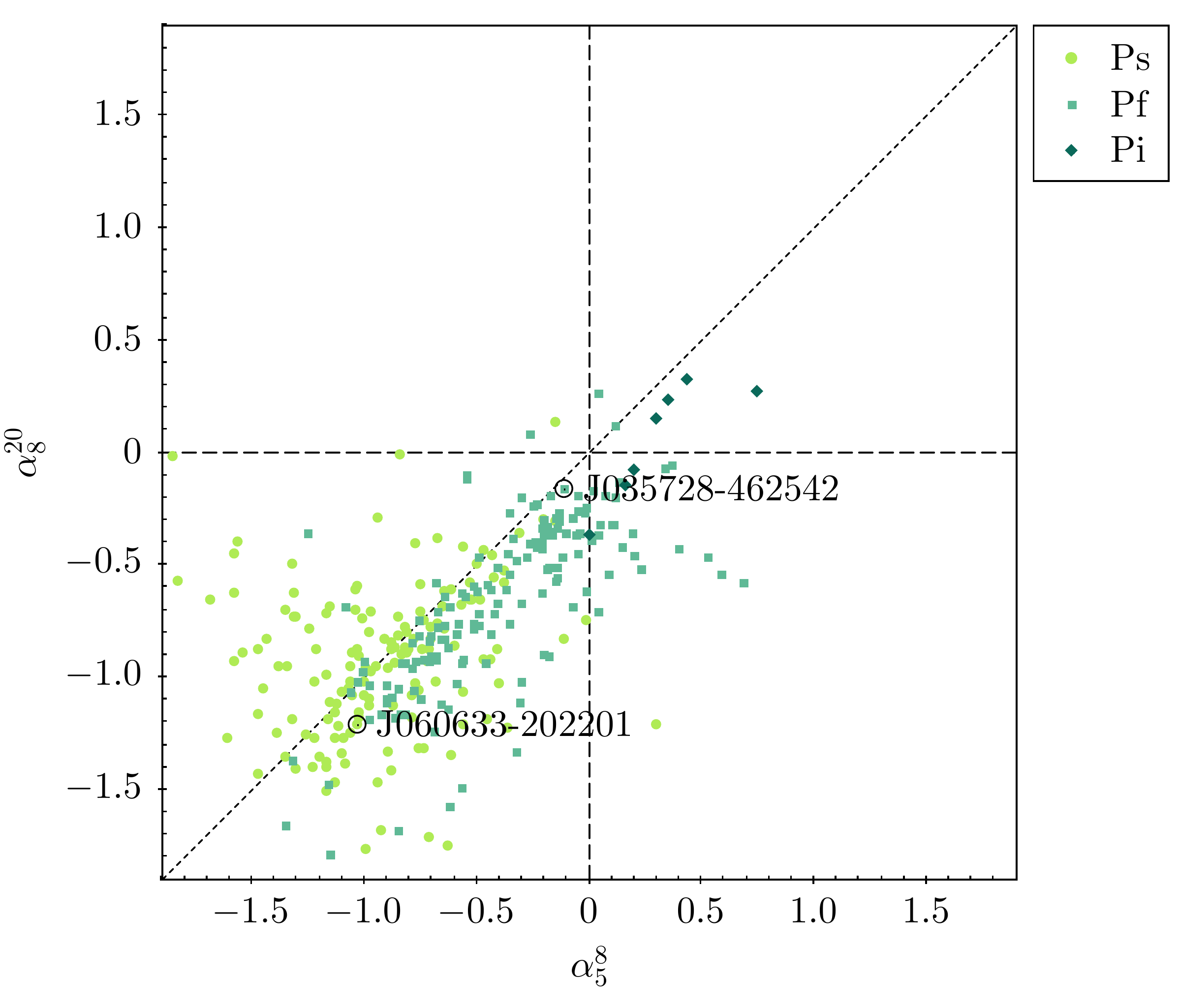}
    \footnotesize (a) Similar to Fig.~\ref{fig:alpha5_8-alpha8_20-fitClass_U}a, plotting the peaked sources from our good fit quality sample. Lime green circles: Ps. Turquoise squares: Pf. Dark green diamonds: Pi. Two sources are beyond axis ranges.
  \end{minipage}

  \begin{minipage}[b]{0.41\columnwidth}
    \centering
    \includegraphics[width=\columnwidth]{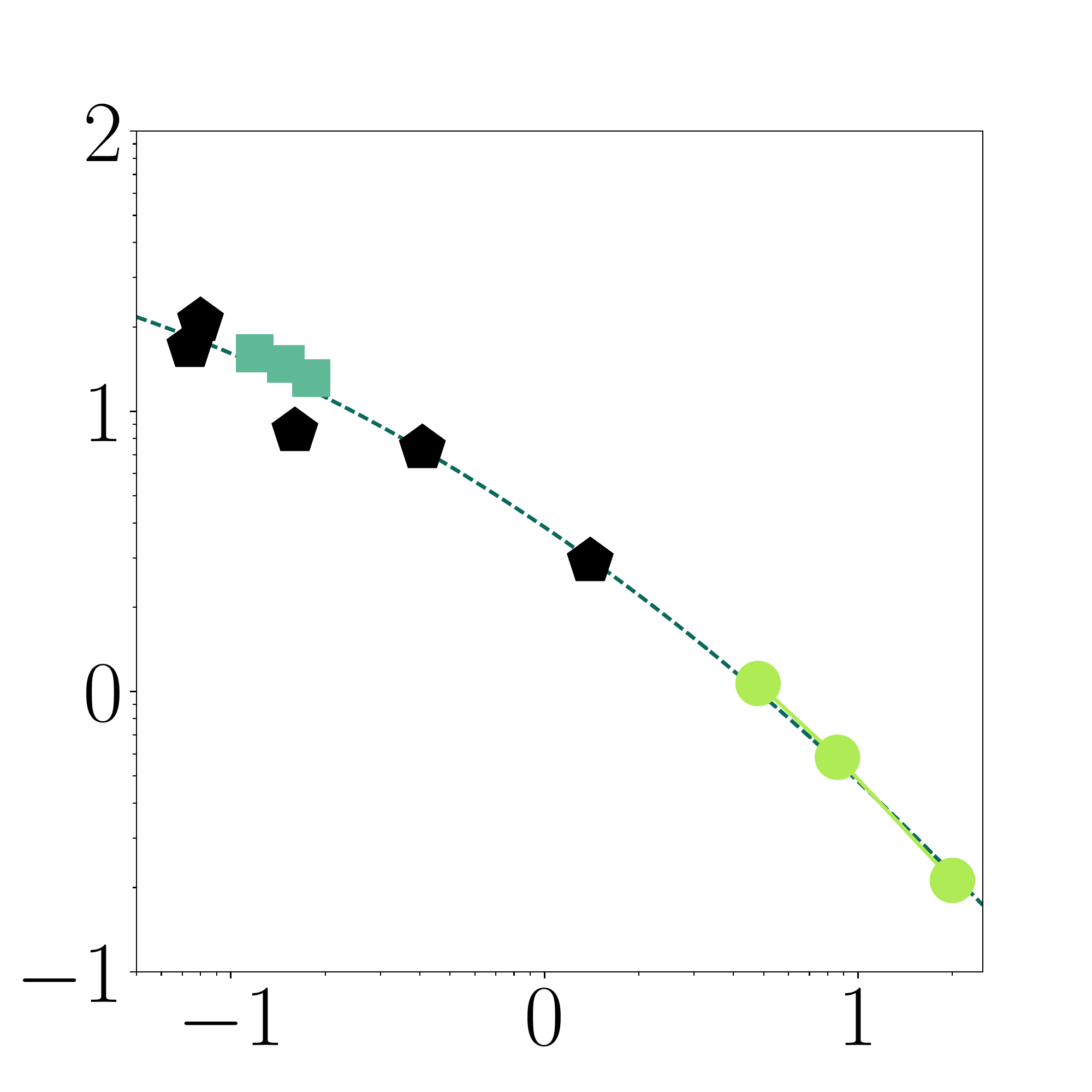}
    \footnotesize (b) J060633-202201
  \end{minipage}%
  \begin{minipage}[b]{0.41\columnwidth}
    \centering
     \includegraphics[width=\columnwidth]{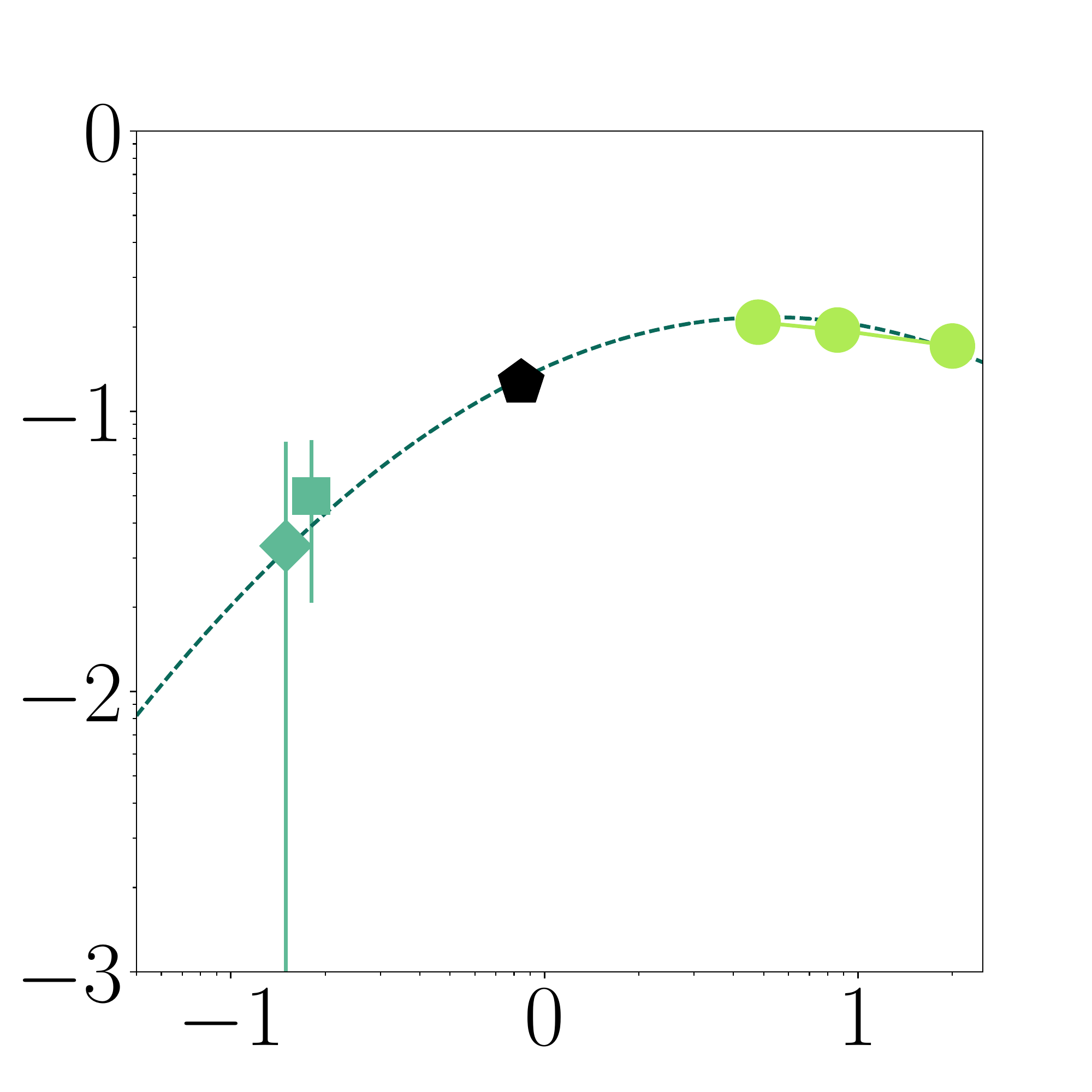}
    \footnotesize (c) J035728-462542
  \end{minipage}%

  \caption{Similar to Fig.~\ref{fig:alpha5_8-alpha8_20-fitClass_L}.}
  \label{fig:alpha5_8-alpha8_20-fitClass_P}
\end{figure}

\subsubsection{Negative curvature}

Fig.~\ref{fig:alpha5_8-alpha8_20-fitClass_P} takes all sources with negative curvature from Fig.~\ref{fig:alpha-phi-fitQual_g} and plots their location by AT20G spectral index data and style and colour by our RSCs. The majority of sources are below the diagonal, however in the steep quadrant we see a mix of Ps and Pf sources. There is nothing unique about the high-frequency spectral indices which allows us to separate those which turn over and peak (or come close, like Fig.~\ref{fig:alpha5_8-alpha8_20-fitClass_P}b) from those which continue to be steep at low frequencies (like Fig.~\ref{fig:alpha5_8-alpha8_20-fitClass_U}b). Only by considering the low-frequency data points can we separate the lobe-emission-dominated steep sources from the CSS sources peaking below \SI{1}{\giga\hertz}.

Fig.~\ref{fig:alpha5_8-alpha8_20-fitClass_P}c shows a source which appears flat at high frequencies, and reveals how such a source may in fact be a GPS source otherwise easy to miss without considering the entire frequency domain.

\subsubsection{Summary}

\begin{figure*}
  \centering
  \includegraphics[width = 1.5\columnwidth]{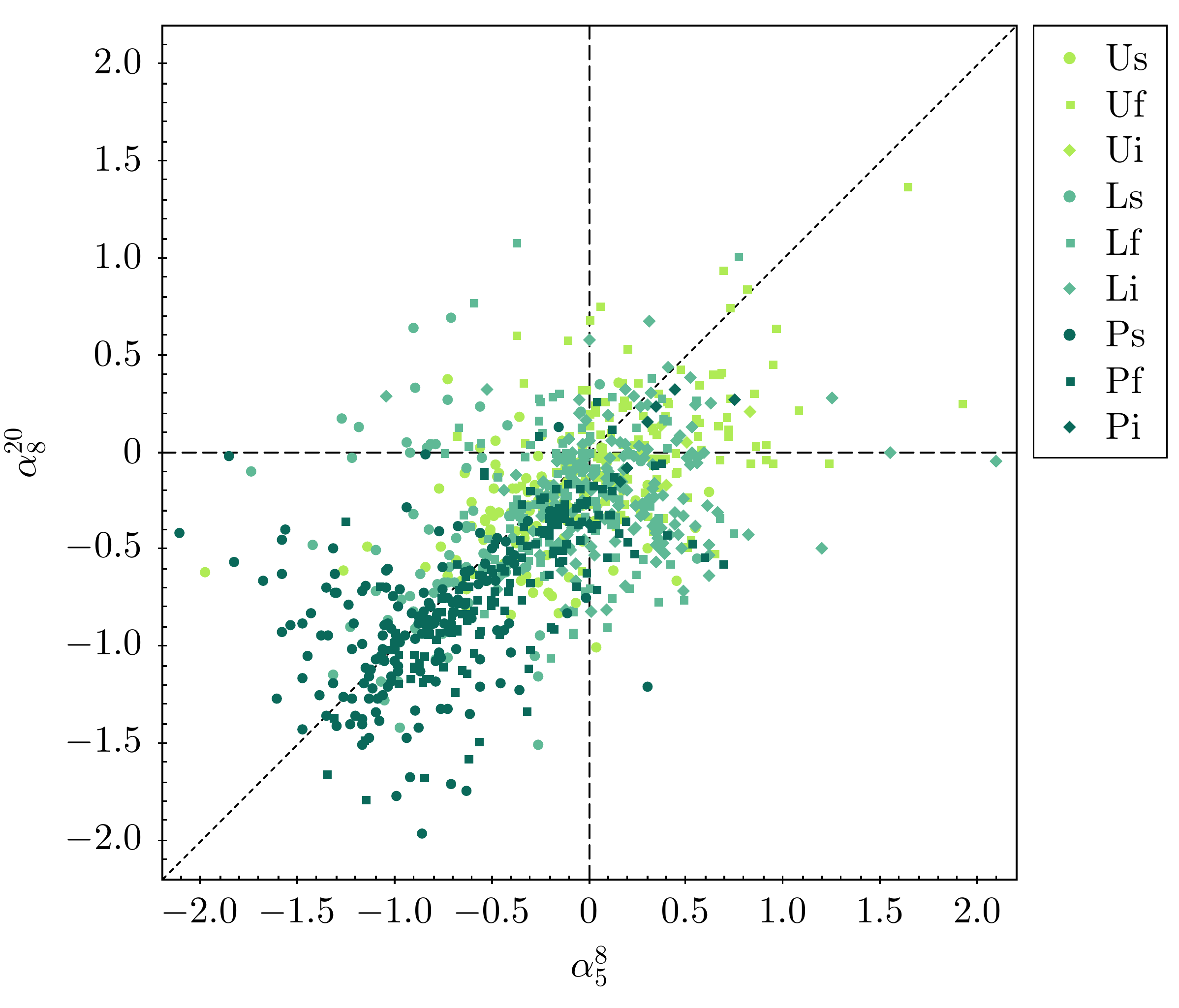}
  \caption{RCC diagram from \SIrange{5}{8}{\giga\hertz} and \SIrange{8}{20}{\giga\hertz} using our good fit quality sample, differentiating superclasses by colour and subclasses by shape. Lime green: Upturned (Fig.~\ref{fig:alpha5_8-alpha8_20-fitClass_U}). Turquoise: Linear (Fig.~\ref{fig:alpha5_8-alpha8_20-fitClass_L}). Dark green: Peaked (Fig.~\ref{fig:alpha5_8-alpha8_20-fitClass_P}). Circles: Steep. Squares: Flat. Diamonds: Inverted. One source is beyond axis ranges.}
  \label{fig:alpha5_8-alpha8_20-fitClass_ULP}
\end{figure*}

It is important to keep in mind that Fig.~\ref{fig:alpha5_8-alpha8_20-fitClass_U} to~\ref{fig:alpha5_8-alpha8_20-fitClass_P} are only subsets of a single RCC diagram, split apart using our broad-frequency RSCs for finer analysis, which is information unavailable using AT20G alone. If we combine them, we form Fig.~\ref{fig:alpha5_8-alpha8_20-fitClass_ULP}, where each of Fig.~\ref{fig:alpha5_8-alpha8_20-fitClass_U} to~\ref{fig:alpha5_8-alpha8_20-fitClass_P}, is represented by a different colour, allowing us to generally compare the superclasses. We notice that the upturned sources cluster around the centre of the diagram, the peaked sources to the lower-left, and the linear sources between those two. This reflects the selection biases brought about by applying RSCs using low-frequency data, however it makes clear the extent to which sources of all types can be entwined throughout the RCC diagram.

\subsection{Comparison to broad frequency}
So far we have demonstrated that the RSC system is superior to methods which have data selected over a narrower frequency range, which is an unfair comparison that serves mostly to point out the large variation possible in SEDs over a wider frequency range. A more informative comparison is to the \spindex[1]{0.1}-\spindex[20]{1} RCC diagram, formed by taking the spectral index between \SIlist{0.18; 1.4}{\giga\hertz} (or upper bound \SI{0.843}{\giga\hertz} if SUMSS is provided instead of NVSS) and the spectral index between \SIlist{1.4; 20}{\giga\hertz} (or lower bound \SI{0.843}{\giga\hertz}). This is about as large a domain of frequencies as a spectral index can cover under the assumption of constancy between points, and creates the sort of RCC diagram which another study combining AT20G and MWACS data might have used. As it happens, it appears to be somewhat effective.

\begin{figure*}
  \centering
  \begin{minipage}[c]{0.7\textwidth}
    \includegraphics[width=\textwidth]{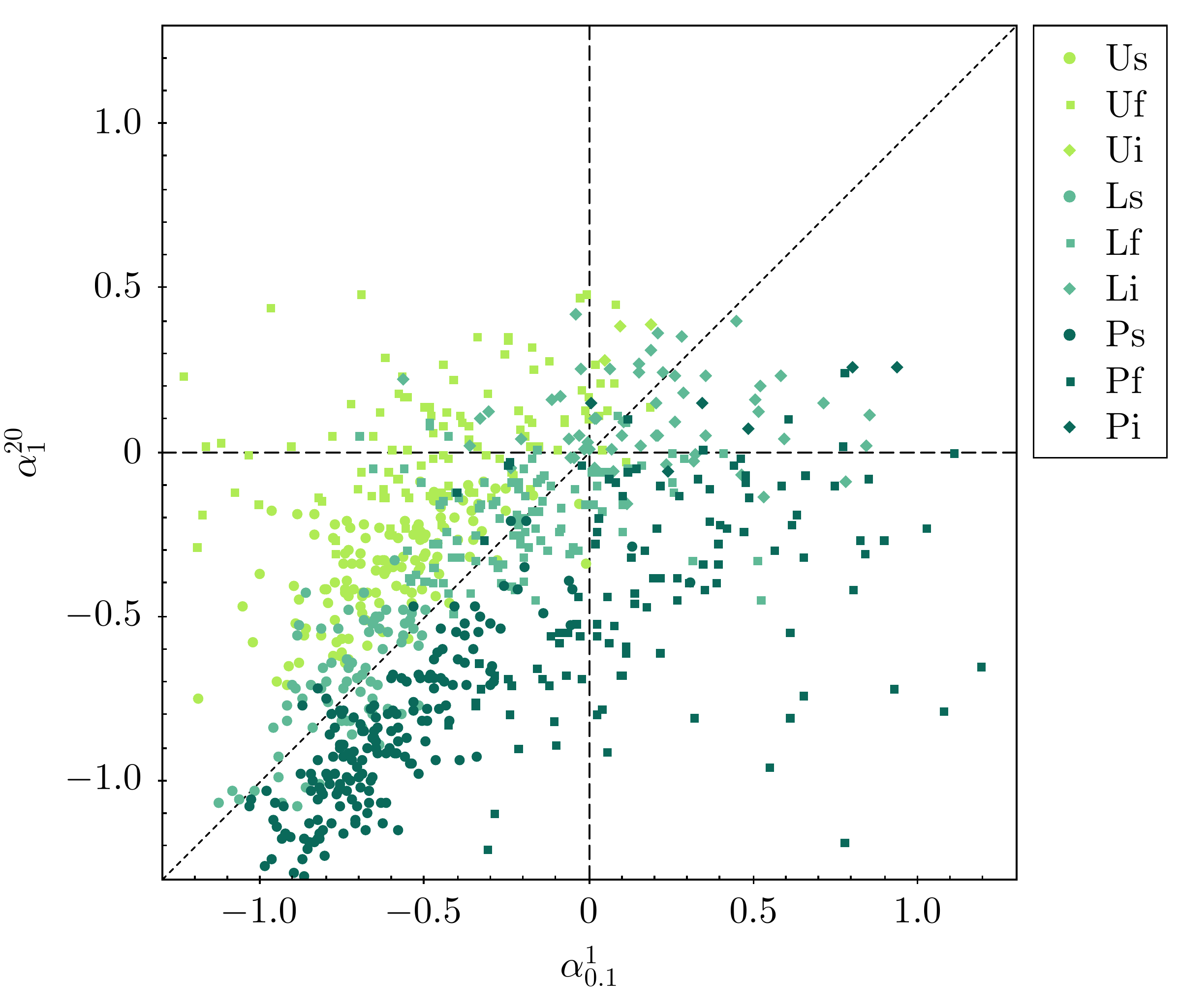}
  \end{minipage}\hfill
  \begin{minipage}[c]{0.28\textwidth}
    \footnotesize (a) Superclasses by colour, subclasses by shape -- Lime green: Upturned. Turquoise: Linear. Dark green: Peaked. Circles: Steep. Squares: Flat. Diamonds: Inverted.
  \end{minipage}
  \\~\\
  \begin{minipage}[c]{0.7\textwidth}
    \includegraphics[width=\textwidth]{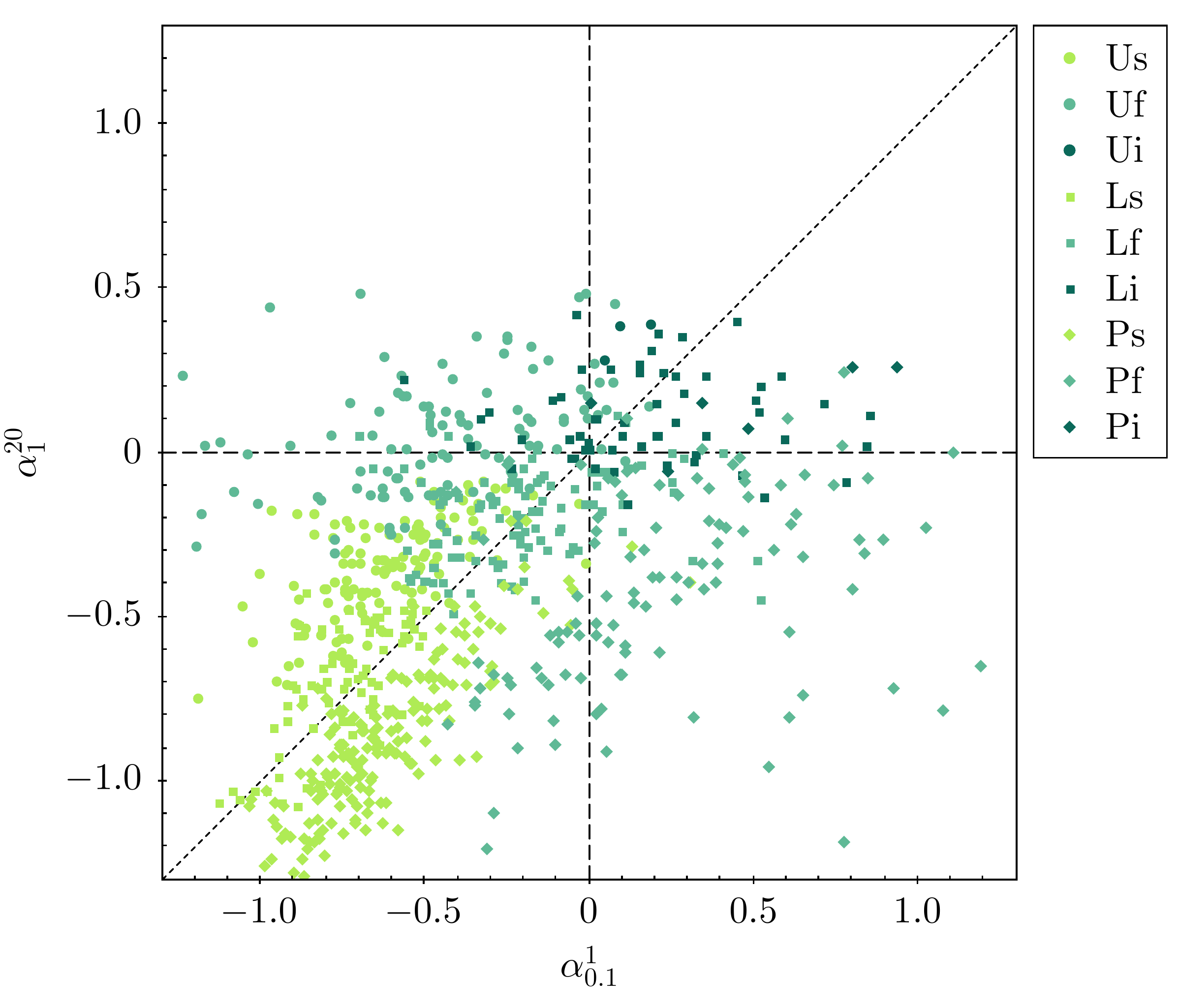}
  \end{minipage}\hfill
  \begin{minipage}[c]{0.28\textwidth}
    \footnotesize (b) Superclasses by shape, subclasses by colour -- Lime green: Steep. Turquoise: Flat. Dark green: Inverted. Circles: Upturned. Squares: Linear. Diamonds: Peaked.
  \end{minipage}
  \caption{RCC diagram from \roughly{0.1}~to \roughly\SI{1}{\giga\hertz} and \roughly\SIrange{1}{20}{\giga\hertz}, using our good fit quality sample, differentiating the super- or subclasses by colour or shape. Nine sources are beyond axis ranges. Sources with no \SI{180}{\mega\hertz} detection are not shown as the lower spectral index cannot be calculated.}
  \label{fig:alpha01_1-alpha1_20-fitClass}
\end{figure*}

From Fig.~\ref{fig:alpha01_1-alpha1_20-fitClass}a it is clear that the RSC system is identifying real trends in the sources. As determined by superclass, upturned sources generally appear above the RCC diagonal, linear sources along it, and peaked sources below it. In Fig.~\ref{fig:alpha01_1-alpha1_20-fitClass}b we see that subclass steep sources are predominantly in the steep quadrant (though more obviously constrained below \(\alpha = -0.5\)), flat sources either near the origin or in the upturned and peaked quadrants, and inverted sources in the inverted quadrant.

There is still a degree of scatter obvious in the classes as they appear in Fig.~\ref{fig:alpha01_1-alpha1_20-fitClass}, which indicates that sources near the edges of RCC quadrants are mischaracterised compared to the \(\alpha\)-\(\phi\) diagram. It is especially evident that picking the `truly linear' sources off an RCC diagram is difficult as it is unclear how far from the diagonal a source is allowed to be while still having a sufficiently linear spectral shape. The RSC definition of linear sources, identified by the curvature term being consistent with zero, is very rigorous, and it is notable that sources which have linear spectra can be quite far from the diagonal in the steep quadrant of an RCC diagram.

It could be suggested that the \(\alpha\)-\(\phi\) diagram is unnecessary as an RCC diagram covering the same broad range of frequencies seems to have success at separating the sources. It is apparent that the radio spectral classification populations are fairly consistent with the quadrants of the RCC diagram. However, we posit that the rigorous definitions of RSC boundaries using $\phi$ and $\sub{\nu}{turn}$ are superior to the fuzzy definition of RCC boundaries using the $\alpha = 0$ quadrant separators. Furthermore, in aid of attaining a single spectral index for linear sources that can be easily read from the plot, an \(\alpha\)-\(\phi\) diagram is clearly more suited than an RCC diagram.

\subsection{Drawbacks}
The most troublesome property of the \(\alpha\)-\(\phi\) diagram is simply that in modern astrophysics, the literature places great focus on spectral indices across small and specific frequency domains, which a standard RCC diagram can work with effectively but an \(\alpha\)-\(\phi\) diagram cannot. Except for linear sources, the \(\alpha\)-\(\phi\) diagram cannot provide a simple view of the spectral index across a small range due to its requirement of a quadratic fit to quantify curvature.

Arguably this is not a fault of the \(\alpha\)-\(\phi\) diagram, but rather the notion in astrophysics that despite a source showing obvious and significant curvature, the spectral index between two points along that curve is meaningful. Considering \percentage{64} of sources are revealed to have spectral curvature in the \(\alpha\)-\(\phi\) diagram, we suggest that our method is a superior descriptor of the intrinsic SEDs of these sources.

\section{Conclusions}
\label{sec:conclusion}

We have created a new source catalogue, AT20G-MWACS, by building on the solid high-frequency foundation of the AT20G survey and combining it with the results of a fresh run of source finding algorithms on the original MWACS images. The catalogue covers an area of \SI{5760}{\sqdegree} and contains 1285 sources.

By fitting models to the data, we have removed the effect of troublesome variations in flux density that can impact attempts to classify sources by spectral index alone. The \(\alpha\)-\(\phi\) diagram, plotting the fitted linear and curvature terms against each other, offers a new visualisation technique for understanding source populations, and with the inherent goodness of fit quantifiers from the modelling process we can focus on only those sources which are appropriate for these models. We say that 994 of the 1285~sources have good fit quality for a linear or quadratic model.

We have split the \(\alpha\)-\(\phi\) diagram into nine classes, and proven these radio spectral classes can be used to conveniently describe source populations. The boundaries between classes are well-defined, unlike the boundaries which must be used when reading sources from a radio colour-colour diagram alone. We have shown the \(\alpha\)-\(\phi\) diagram to be superior in readability and population identification to radio colour-colour diagrams across both narrow and broad frequency ranges. It is clear that proper characterisation of the spectral form of a source requires a broad frequency domain, and that predictions made from narrow-band measurements alone are extremely insufficient.

Future work might begin with constructing \(\alpha\)-\(\phi\) diagrams for other similarly broad-frequency source sets which could be used to determine how well the process scales to other frequency domains. The first release from the Galactic and Extragalactic All-Sky MWA survey \cite[GLEAM;][]{wayth2015,hurley-walker2016} may be of particular use, most notably for its higher angular resolution and sampling rate across a significantly broader frequency domain than MWACS. As GLEAM covers the whole southern sky, the entire AT20G catalogue can be cross-matched to potentially create one of the largest, broadest, and deepest radio source catalogues to date.

\section*{Acknowledgements}
V.M.H. recognises the ICRAR Summer Studentship Program and the Pawsey Supercomputing Centre Summer Internship Program for providing travel, accommodation, and support funding which made this research possible.

Colour palettes adapted for colour blindness sourced from research by Martin Krzywinski (\href{http://mkweb.bcgsc.ca/colorblind}{\texttt{http://mkweb.bcgsc.ca/colorblind}}).

We thank an anonymous referee for their kind comments and helpful suggestions for improvements to the manuscript.

\setlength{\labelwidth}{0pt}

\label{lastpage}


\begin{thebibliography}{}

\bibitem[\protect\citeauthoryear{Bennett}{1962}]{Bennett:62}
Bennett~A.~S., 1962, MNRAS, 125, 75

\bibitem[\protect\citeauthoryear{Bolton~J.~G.}{1969}]{bolton1969}
Bolton~J.~G., 1969, AJ, 74, 131

\bibitem[\protect\citeauthoryear{Burnham \& Anderson}{2002}]{Burnham2002}
Burnham~K.~P. \& Anderson~D.~R., 2002, Model Selection and Multimodel Inference: A Practical Information-Theoretical Approach. 2nd ed. New York: Springer-Verlag

\bibitem[\protect\citeauthoryear{Callingham et al.}{2015}]{Callingham:15}
Callingham~J.~R., et al., 2015, ApJ, 809, 168

\bibitem[\protect\citeauthoryear{Chhetri et al.}{2013}]{chhetri2013}
Chhetri~R., Ekers~R.~D., Jones~P.~A., Ricci~R., 2013, MNRAS, 434, 956

\bibitem[\protect\citeauthoryear{Clemens et al.}{2010}]{Clemens:10}
Clemens~M.~S., Scaife A., Vega~O., Bressan~A., 2010, MNRAS, 405, 887

\bibitem[\protect\citeauthoryear{Cohen et al.}{2007}]{cohen2007}
Cohen~A.~S., Lane~W.~M., Cotton~W.~D., Kassim~N.~E., Lazio~T.~J.~W., Perley~R.~A., Condon~J.~J., Erickson~W.~C., 2007, AJ, 134, 1245

\bibitem[\protect\citeauthoryear{Condon}{1992}]{Condon:92}
Condon~J.~J., 1992, ARA\&A, 30, 575

\bibitem[\protect\citeauthoryear{Condon et al.}{1998}]{condon1998}
Condon~J.~J., Cotton~W.~D., Greisen~E.~W., Yin~Q.~F., Perley~R.~A.,Taylor~G.~B., Broderick~J.~J., 1998, AJ, 115, 1693

\bibitem[Conway et al.(1963)]{Conway:63}
Conway~R.~G., Kellermann~K.~I., \& Long, R.~J.\ 1963, MNRAS, 125, 261

\bibitem[\protect\citeauthoryear{De Breuck et al.}{2002}]{debreuck2002}
De~Breuck~C., Tang~Y., de Bruyn~A.~G., R{\"o}ttgering~H., van Breugel~W., 2002, A\&A, 394, 59

\bibitem[\protect\citeauthoryear{Franzen et al.}{2014}]{Franzen:14}
Franzen~T.~M.~O., et al., 2014, MNRAS, 439, 1212

\bibitem[\protect\citeauthoryear{Galvin et al.}{2016}]{Galvin:16}
Galvin T.~J., Seymour N., Filipovi{\'c} M.~D., Tothill N.~F.~H., Marvil J., Drouart G., Symeonidis M., Huynh M.~T., 2016, MNRAS, 461, 825

\bibitem[\protect\citeauthoryear{Hancock et al.}{2012}]{hancock2012}
Hancock~P. et al., 2012, MNRAS, 422, 1812

\bibitem[\protect\citeauthoryear{Hurley-Walker et al.}{2014}]{hurley-walker2014}
Hurley-Walker~N. et al., 2014, PASA, 31, e045

\bibitem[\protect\citeauthoryear{Hurley-Walker et al.}{2016}]{hurley-walker2016}
Hurley-Walker~N. et al., 2016, MNRAS, 464/1/1146

\bibitem[Kellermann \& Pauliny-Toth (1969)]{Kellermann:69}
Kellermann~K.~I., Pauliny-Toth~I.~I.~K.\ 1969, ApJL, 155, L71

\bibitem[\protect\citeauthoryear{Kesteven et al.}{1977}]{kesteven1977}
Kesteven~M.~J.~L., 1977, AJ, 82, 541

\bibitem[\protect\citeauthoryear{Large et al.}{1981}]{large1981}
Large~M.~I., Mills~B.~Y., Little~A.~G., Crawford~D.~F., Sutton~J.~M., 1981, MNRAS, 194, 693

\bibitem[\protect\citeauthoryear{Mauch et al.}{2007}]{mauch2007}
Mauch~T., Murphy~T., Buttery~H.~J., Curran~J., Hunstead~R.~W., Piestrzynski~B., Robertson~J.~G., Sadler~E.~M., 2007, VizieR Online Data Catalog, 8081, 0

\bibitem[\protect\citeauthoryear{Murphy et al.}{2007}]{murphy2007}
Murphy~T., Mauch~T., Green~A., Hunstead~R.~W., Piestrzynska~B., Kels~A.~P., Sztajer~P., 2007, MNRAS, 382, 382

\bibitem[\protect\citeauthoryear{Murphy et al.}{2010}]{AT20Gsource}
Murphy~T. et al., 2010, MNRAS, 402, 2403

\bibitem[\protect\citeauthoryear{O'Dea}{1998}]{ODea:98}
O'Dea C.~P., 1998, PASP, 110, 493

\bibitem[\protect\citeauthoryear{Rudnick, Katz-Stone \& Anderson}{1994}] {Rudnick:94}
Rudnick L., Katz-Stone D.~M., Anderson M.~C., 1994, ApJS, 90, 955

\bibitem[\protect\citeauthoryear{Schmitt}{1968}]{Schmidt:68}
Schmitt J.~L., 1968, Natur, 218, 663

\bibitem[\protect\citeauthoryear{Slee}{1995}]{slee1995}
Slee~O.~B., 1995, AuJPh, 48, 143

\bibitem[\protect\citeauthoryear{Wayth et al.}{2015}]{wayth2015}
Wayth~R. et al., PASA, 2015, 32, e025

\bibitem[\protect\citeauthoryear{Williams}{1963}]{Williams:63}
Williams P.~J.~S., 1963, Natur, 200, 56
\end{thebibliography}
\end{document}